\documentclass[aip,pof,amsmath,amssymb,%
reprint,%
]{revtex4-1}

\usepackage{graphicx}
\usepackage{color}
\usepackage{dcolumn}
\usepackage{epstopdf}
\usepackage{bm}
\usepackage{amsthm}
\usepackage{afterpage}

\begin{document}
\preprint{AIP/123-QED}
\newcommand{\figpath}{./}

\newcommand{\solid}{\hbox{---$\!$---}\nobreak}
\newcommand{\solidcirc}{\hbox{---$\circ$---}\nobreak}
\newcommand{\solidcross}{\hbox{---$\times$---}\nobreak}
\newcommand{\solidplus}{\hbox{---$+$---}\nobreak}
\newcommand{\solidsquare}{\hbox{---$\square$---}\nobreak}
\newcommand{\solidcircf}{\hbox{---$\bullet$---}\nobreak}
\newcommand{\solidsquaref}{\hbox{---$\blacksquare$---}\nobreak}
\newcommand{\solidtri}{\hbox{---$\vartriangle$---}\nobreak}
\newcommand{\solidtridown}{\hbox{---$\triangledown$---}\nobreak}
\newcommand{\solidtrif}{\hbox{---$\blacktriangle$---}\nobreak}
\newcommand{\dashed}{\hbox{{--}\,{--}\,{--}\,{--}}}
\newcommand{\chndot}{\hbox{---\,$\cdot$\,---}\nobreak}
\newcommand{\chndotsquare}{\hbox{--\,$\cdot$\,--$\square$--\,$\cdot$\,--}}
\newcommand{\chndottri}{\hbox{--\,$\cdot$\,--$\vartriangle$--\,$\cdot$\,--}\nobreak}
\newcommand{\dashedcirc}{\hbox{{--}\,{--}\,$\circ${--}\,{--}}}
\newcommand{\dashedcircf}{\hbox{{--}\,{--}\,$\bullet${--}\,{--}}}
\newcommand{\dashedsquare}{\hbox{{--}\,{--}\,$\square${--}\,{--}}}
\newcommand{\dashedtri}{\hbox{{--}\,{--}\,$\vartriangle${--}\,{--}}}
\newcommand{\dashedtridown}{\hbox{{--}\,{--}\,$\triangledown${--}\,{--}}}
\newcommand{\itbold}[1]{\textbf{\textit{#1}}}
\newcommand{\divorig}{\div}
\newcommand{\rot}[1]{\nabla \times \textbf{\textit{#1}}}
\renewcommand{\div}[1]{\nabla \cdot \textbf{\textit{#1}}}
\newcommand{\dif}[2]{\dfrac{\textup{d} #1}{\textup{d} #2}}
\newcommand{\ddif}[2]{\dfrac{\textup{d}^2 #1}{\textup{d}{ #2}^2}}
\newcommand{\pdif}[2]{\dfrac{\partial #1}{\partial #2}}
\newcommand{\pddif}[2]{\dfrac{\partial^2 #1}{\partial {#2}^2}}
  \newcommand{\up}{{u}^\prime}
  \newcommand{\uvec}{\itbold{u}}
  \newcommand{\upvec}{\itbold{u}^\prime}
  \newcommand{\Uvec}{\itbold{U}}
  \newcommand{\kvec}{\itbold{k}}
  \newcommand{\qvec}{\itbold{q}}
  \newcommand{\xvec}{\itbold{x}}
  \newcommand{\Hvec}{\itbold{H}}
 
  \newcommand{\omgvec}{\boldsymbol{\omega}}	  
  \newcommand{\Omgvec}{\boldsymbol{\Omega}}	  
  
  \newcommand{\NL}{(\uvec\cdot\nabla)\uvec}
  \newcommand{\NLT}{(\uvec\cdot\nabla)T}

  \newcommand{\urms}{u^{\prime}}
  \newcommand{\vrms}{v^{\prime}}
  \newcommand{\wrms}{w^{\prime}}
  
  \newcommand{\upls}{u^{\prime +}}
  \newcommand{\vpls}{v^{\prime +}}
  \newcommand{\wpls}{w^{\prime +}}

  \newcommand{\uup}{\langle {u}^{2} \rangle}
  \newcommand{\vvp}{\langle {v}^{2} \rangle}
  \newcommand{\wwp}{\langle {w}^{2} \rangle}
  \newcommand{\uvp}{\langle {u} {v} \rangle}

  \newcommand{\diss}{\langle \varepsilon \rangle}
  \newcommand{\tdiss}{\varepsilon_t }
  
  \newcommand{\II}{$I\hspace{-.1em}I$}
  \newcommand{\III}{$I\hspace{-.1em}I\hspace{-.1em}I$}
  \newcommand{\IV}{$I\hspace{-.1em}V$}
  
\def\dd{{\, \rm{d}}}
\def\dr{{\rm{d}}}
\def\Dr{{\rm{D}}}
\def\bra{\langle}
\def\ket{\rangle}
\def\p{\partial}
\def\ra{\rightarrow}
\def\beq{\begin{equation}}
\def\eeq{\end{equation}}
\def\la{\label}
\def\r#1{(\ref{#1})}
\def\ii{{\rm i}}
\def\ee{{\rm e}}
\def\hu{\widehat{u}}
\def\hv{\widehat{v}}
\def\hw{\widehat{w}}
\def\hp{\widehat{p}}
\def\homega{\widehat{\omega}}
%

\title[DNS of statistically stationary HST]%
{Direct numerical simulation of statistically stationary and homogeneous shear turbulence
and its relation to other shear flows}

\author{Atsushi Sekimoto}
\email{sekimoto@torroja.dmt.upm.es, a.sekki@gmail.com}
\affiliation{School of Aeronautics, Universidad Polit\'ecnica de Madrid, 28040 Madrid, Spain}
\author{Siwei Dong}
\affiliation{School of Aeronautics, Universidad Polit\'ecnica de Madrid, 28040 Madrid, Spain}
\author{Javier Jim\'enez}
\email{jimenez@torroja.dmt.upm.es}
\affiliation{School of Aeronautics, Universidad Polit\'ecnica de Madrid, 28040 Madrid, Spain}

\date{\today}

\begin{abstract}
Statistically stationary and homogeneous shear turbulence (SS-HST) is investigated by
means of a new direct numerical simulation code, spectral in the two horizontal directions
and compact-finite-differences in the direction of the shear. No remeshing is used to
impose the shear-periodic boundary condition. The influence of the geometry of the
computational box is explored. Since HST has no characteristic outer length scale and
tends to fill the computational domain, long-term simulations of HST are `minimal' in
the sense of containing on average only a few large-scale structures. It is found
that the main limit is the spanwise box width, $L_z$, which sets the length and
velocity scales of the turbulence, and that the two other box dimensions should be
sufficiently large $(L_x\gtrsim 2L_z$, $L_y \gtrsim L_z$) to prevent other directions
to be constrained as well. It is also found that very long boxes, $L_x \gtrsim 2
L_y$, couple with the passing period of the shear-periodic boundary condition, and
develop strong unphysical linearized bursts. Within those limits, the flow shows
interesting similarities and differences with other shear flows, and in particular
with the logarithmic layer of wall-bounded turbulence. They are explored in some
detail. They include a self-sustaining process for large-scale streaks and
quasi-periodic bursting. The bursting time scale is approximately universal, $\sim 20
S^{-1}$, and the availability of two different bursting systems allows the growth of
the bursts to be related with some confidence to the shearing of initially isotropic
turbulence. It is concluded that SS-HST, conducted within the proper
computational parameters, is a very promising system to study shear turbulence in
general.
\end{abstract}

\pacs{Valid PACS appear here}
\maketitle

\section{Introduction}

The simulation of ever higher Reynolds numbers is a common practice in modern turbulence
research, mainly to study the multi-scale nature of the energy cascade and similar
phenomena. A basic property of these processes is that they are nonlinear and chaotic,
which makes their theoretical analysis difficult. Direct numerical simulations offer
a promising tool for their study because they provide exceptionally rich data sets.
They have also been moving recently into the same range of Reynolds numbers as most
experiments, especially when similar levels of observability are compared.

On the other hand, simulations are not without problems, some of which they share
with experiments. One of those problems is forcing. Turbulence is dissipative, and
unforced turbulence quickly decays. The highest resolution simulations available,
with Taylor-microscale Reynolds number $Re_\lambda \sim O(1000)$, are nominally
isotropic flows with artificial forcing in a few low wavenumbers,~\cite{kaneda03} and
it is difficult to judge how far into the cascade the effect of the forcing extends.
The forcing of wall-bounded shear flows is well characterized and corresponds to
physically realizable situations, and they have often been used as alternatives to
study high-Reynolds number turbulence. However, a consequence of their more natural
large scales is that their cascade range is much shorter than in the isotropic boxes.
Current simulations only reach $Re_\lambda \sim 200$ (see
Refs.~\onlinecite{LozanoJimenez2014pof,BernardiniPirozzoli2014}). They are also
inhomogeneous, and it is difficult to distinguish which properties are due to the
inhomogeneity, to the presence of the wall, or to inertial turbulence itself.

A compromise between the two limits is homogeneous shear turbulence (HST), which shares
the natural energy-generation mechanism of shear flows with the simplicity of homogeneity.
This flow is believed not to have an asymptotic statistically stationary state, but it is
tempting to use it as a proxy for general shear turbulence in which high Reynolds numbers
can be reached without the complications of wall-bounded flows. Up to recently, HST has
mostly been used to study the generation of turbulent fluctuations during the initial
shearing of isotropic turbulence, where one of the classic challenges is to
determine the growth rate with a view to developing turbulence models, avoiding as far as possible artifacts due to numerics and  initial conditions. A question
closer to our interest in this paper is whether some aspects of HST can be used as models
for other shear flows. For example, Rogers and Moin~\cite{RogersMoin1985} showed typical
`hairpin' structures under shear rates comparable to those in the logarithmic layer of
wall-bounded flows, and Lee {\it et al.}~\cite{LeeKimMoin1990} found that higher shear
rates, comparable to those in the buffer layer, result in structures similar to near-wall
velocity streaks.~\cite{Jimenez2013nearwall} Those structures are known to play crucial
roles in transition and in maintaining shear-induced
turbulence,~\cite{JimenezMoin1991,jim94,HamiltonKimWaleffe1995,Waleffe1997} and
Kida and Tanaka~\cite{KidaTanaka1994} proposed a generation mechanism for the streamwise
vortices in transient HST that recalls those believed to be active in wall turbulence.

Ideal HST in unbounded domains grows indefinitely, both in intensity and length
scale.~\cite{RogersMoin1985,TavoularisKarnik1989,IsazaCollins2009} During the initial
stages of shearing an isotropic turbulent flow, linear effects result in algebraic growth
of the turbulent kinetic energy, which is later transformed to exponential due to
nonlinearity.~\cite{CambonScott1999} 
These simulations are typically discontinued as the growing length scale of the sheared
turbulence approaches the size of the computational box, but Pumir~\cite{Pumir1996}
extended the simulation to longer times and reached a statistically stationary state
(SS-HST) in which the largest-scale motion is constrained by the computational box and
undergoes a succession of growth and decay of the kinetic energy and of the enstrophy
reminiscent of the bursting in wall-bounded flows,~\cite{jim05} suggesting that bursting
is a common feature of shear-induced turbulence, not restricted to wall-bounded
situations. In fact, previous investigations of SS-HST have suggested that the
growth phase of bursts is qualitatively similar to the initial shearing of isotropic
turbulence.~\cite{Pumir1996,Gualtieri2002}

It should be emphasized that the main goal of this paper is not to
investigate the properties of unbounded HST, which is difficult to implement both
experimentally and numerically for the reasons explained above. Simulations in a finite
box introduce a length scale that, without interfering with homogeneity, is incompatible
with an unbounded flow. It is precisely in the effect of this extraneous length scale,
which may make simulations more similar to flows in which a length scale is enforced by
the wall, that we are more interested. As first pointed out in Ref.
\onlinecite{Pumir1996}, this is what SS-HST provides.

In this paper we present a numerical code optimized to perform the long simulations
required to study SS-HST. We characterize the code and the influence of the
numerical parameters on the physics, and draw preliminary conclusions about the
physics itself. Since it will be seen that bursting times are $O(20 S^{-1})$, where
$S$ is the mean velocity gradient, the required simulation times are of the order of
several hundreds shear times. The most commonly used code for HST is due to
Rogallo,~\cite{Rogallo1981} and involves remeshing every few shear times. Some
enstrophy is lost in each remeshing \cite{LeeKimMoin1990}, and the concern about the
cumulative effect of these losses has lead to the search for improved simulation
schemes. Schumacher {\it et al.}~\cite{SchumacherEckhardt2000}, and
Horiuchi~\cite{Horiuchi2011} introduced artificial body forces to drive a mean shear
gradient between stress-free surfaces, but this creates thin layers near the
surfaces, and the impermeability condition prevents large-scale motions from
developing. A different strategy for avoiding discrete remeshings in fully spectral
codes is that of Brucker {\it et al.},~\cite{BruckerIsazaCollins2007} who used an
especially developed Fourier transform 
essentially equivalent to remeshing at each time step.
    
Another approach to avoid periodic remeshing was pioneered by Baron~\cite{Baron1982} and
later by Schumann~\cite{Schumann1985} and Gerz {\it et
al.},~\cite{GerzSchumannElghobashi1988} who used a `shear-periodic' boundary condition in
which periodicity is enforced between shifting points of the upper and bottom boundaries
of the computational box by a central-finite-differences scheme. Similar boundary
conditions have been used in simulations of astrophysical
disks~\cite{BalbusHawley1998,Johansen2009zonal,SalhiJacobitzSchneiderCambon2014} under the
name of `shearing-boxes'. The code in this paper belongs to this family, but uses
higher-order approximations and a vorticity representation similar to that in
Ref.~\onlinecite{KimMoinMoser1987}.

A substantial part of the present paper is devoted to the choice of the dimensions of the
computational box. The previous discussion about length scales suggests that this
choice should influence the properties of the resulting SS-HST, but the matter has
seldom been addressed in detail. We also spend some effort comparing the bursts
in SS-HST with those in the logarithmic layer, and with the initial shearing of
isotropic turbulence.

The organization of this paper is as follows. The numerical technique and the shear-periodic
boundary condition is introduced and analyzed in \S\ref{sec:num}. The effect of the
computational domain is studied in \S\ref{sec:boxes}, resulting in the identification of an
acceptable range of aspect ratios in which the flow is as free as possible of computational
artifacts, as well as in the determination of the relevant length and velocity scales.
Sec.~\ref{sec:ssp} contains preliminary comparisons of the results of our simulations with
other shear flows, with emphasis on the logarithmic layer of wall-bounded turbulence.
Conclusions are offered in \S\ref{sec:conc}. Several appendices, included as supplementary
material~\cite{Supplemental1}, contain details of the numerical implementation, and
validations of the code.

\section{Numerical method}\la{sec:num}
\subsection{Governing equations and integration algorithm}\la{sec:numalg}

The velocity and vorticity of an uniform incompressible shear flow are expressed as
perturbations, $\widetilde{\uvec} = \Uvec+ \uvec $ and $\widetilde{\omgvec} = \nabla \times
\widetilde{\uvec} = \Omgvec + \omgvec$, with respect to their ensemble averages
(implemented here as spatial and temporal averaging), $ \Uvec = \langle
\widetilde{\uvec} \rangle = (S y, 0, 0) $ and $ \Omgvec = \langle \widetilde{\omgvec}
\rangle =(0,0, -S )$. The streamwise, vertical, and spanwise directions are, respectively,
$(x, y, z)$ and the corresponding velocity components are $(u,v,w)$. We will occasionally
use time-dependent averages over the computational box, $\langle \cdot \rangle_V$, and over
individual horizontal planes, $\langle \cdot \rangle_{xz}$. Capital letters, such as $U$,
will be mostly reserved for mean values, and primes, such as $u'$, for standard deviations.

The Navier--Stokes equations of motion, including continuity, are reduced to
evolution equations for $\phi =\nabla^2 v$ and $\omega_y$,~\cite{KimMoinMoser1987} with
the advection by the mean flow explicitly separated,
\beq
\pdif{\omega_y}{t} + Sy\pdif{\omega_y}{x} = h_g + \nu \nabla^2 \omega_y,
\qquad
\pdif{\phi }{t} + Sy\pdif{\phi }{x} = h_v + \nu \nabla^2 \phi, 
\label{eq:vorphi}
\eeq
where $\nu$ is the kinematic viscosity. Defining $\Hvec = \uvec \times \omgvec$,
\beq
h_g \equiv  \pdif{H_x}{z} - \pdif{H_z}{x} - S\pdif{v}{z}, 
\qquad
h_v \equiv - \pdif{}{y} \left( \pdif{H_x}{x} + \pdif{H_z}{z} \right) 
            + \left( \pddif{}{x} + \pddif{}{z} \right) H_y.
\la{eq:hgv}
\eeq
In addition, the governing equation for $\langle \uvec \rangle_{xz} $ are
\beq
\pdif{\langle u \rangle_{xz}}{t} = 
-\pdif{\langle u v\rangle_{xz}}{y} + \nu \pddif{\langle u \rangle_{xz}}{y},
\qquad
\pdif{\langle w \rangle_{xz}}{t} = 
-\pdif{\langle w v\rangle_{xz}}{y} + \nu \pddif{\langle w \rangle_{xz}}{y}. 
\label{eq:u00w00} 
\eeq
Continuity requires that the mean vertical velocity should be independent of $y$, and
it is explicitly set to $\bra v\ket_{xz}=0$. We consider a periodic computational
domain, $0\le x \le L_x$, $ 0 \le z \le L_z$, and use two-dimensional
Fourier-expansions with 3/2 dealiasing in those directions. The grid dimensions,
$N_x$ and $N_z$, and their corresponding resolutions, $\Delta x = L_x/N_x$ and
$\Delta z = L_z/N_z$, will be expressed in terms of Fourier modes. The domain is
shear-periodic in $ -L_y/2 \le y \le L_y/2$, as explained in the next section. The $y$ direction is
not dealiased, and $\Delta y = L_y/ N_y$. Its discretization is spectral compact
finite differences with a seven-point stencil on a uniform grid. Its coefficients are
chosen for sixth-order consistency, and for spectral-like behavior at $k_y \Delta
y/\pi = 0.5, 0.7, 0.9 $ (see Ref.~\onlinecite{Lele1992}).

Time stepping is third-order explicit Runge--Kutta~\cite{Spalart1991} modified by an
integrating factor for the mean-flow advection (see appendix A in \cite{Supplemental1}).
This uncouples the CFL condition from the mean flow, and is especially helpful in tall
computational domains. The resulting time step is
 \beq\label{eq:cfl}
   \Delta t \le \mathrm{CFL~min}\!\left(\frac{\Delta x}{\pi|u|},
     \frac{\Delta y}{2.85 |v|},~ \frac{\Delta z}{\pi|w|},
     \frac{(\Delta x)^2}{\pi^2\nu},~ \frac{(\Delta y)^2}{9\nu},
     \frac{(\Delta z)^2}{\pi^2\nu}
     \right),
 \eeq
where the denominators 2.85 and 9 in Eq.~(\ref{eq:cfl}) are the maximum eigenvalues of our
finite differences for the first and second derivatives. An implicit viscous part of
the time stepper is not required because only the advective CFL matters
in turbulent flows with uniform grids of the order of the Kolmogorov viscous scale.
Our simulations use $\mbox{CFL}=0.5 - 0.8$.

The problem has three dimensionless parameters that are chosen to be two aspect ratios of
the computational box, $A_{xz} = L_x/L_z$ and $A_{yz} = L_y/L_z$, and the Reynolds number
based on the box width, $Re_z = SL_z^2/\nu$.

We will also use the Taylor-microscale Reynolds number,  
\beq
Re_\lambda  \equiv \left(\frac{q^2}{3}\right)^{1/2}\frac{\lambda}{\nu} 
               = \left( \frac{5 }{3 \nu \diss} \right)^{1/2} q^2,
\la{eq:relambda}
\eeq
where $\diss = \nu \langle |\omgvec|^2\rangle$ is the dissipation rate of the
fluctuating energy, $q^2=\bra u_i u_i \ket$ is twice the kinetic energy per unit mass, and
repeated indices imply summation. The Corrsin shear parameter \cite{Corrsin1958} is
$S^{\ast} = S q^2/\diss$, and the Kolmogorov viscous length is
$\eta=(\nu^3/\diss)^{1/4}$.

\subsection{The shear-periodic boundary condition}\label{sec:shear-periodic-bc}
\begin{figure}[tbp]
  \centering
   \includegraphics[width=0.80\linewidth,clip]{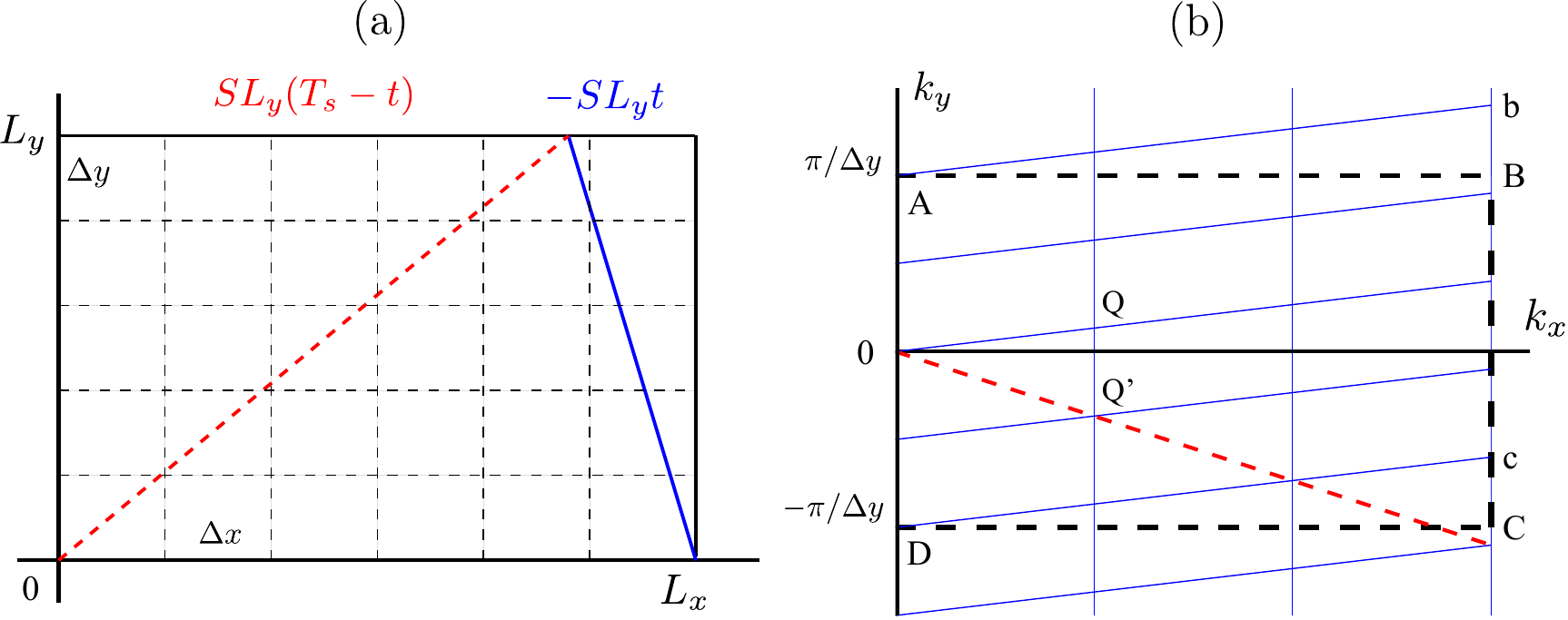}
\caption{Sketch of the shear-periodic boundary condition, as discussed in the text.
(a) Physical space. (b) Fourier space.
}\label{fig:figgrid}
\end{figure} 

The boundary condition used in this study is that the velocity is periodic between
pairs of points in the top and bottom boundaries of the computational box, which are
shifted in time by the mean
shear.~\cite{Baron1982,Schumann1985,GerzSchumannElghobashi1988,BalbusHawley1998} For
a generic fluctuation $g$,
\beq
g(t, x, y, z) = g[t, x + m S t  L_y + l L_x, y + m L_y , z + n L_z], 
\la{eq:bc1}
\eeq
where $l,m$ and $n$ are integers. 
In terms of the spectral coefficients of the expansion,
\beq
g(t, x,y,z) = \sum_{k_x} \sum_{k_z} \widehat{g}(t,k_x,y,k_z) \exp[\ii (k_x x + k_z z)],
\label{eq:shear-periodic-fou}
\eeq
the boundary condition becomes 
\beq
\widehat{g}(t, k_x, y, k_z) = \widehat{g}(t, k_x, y + m L_y, k_z) \exp[\ii k_x m S t L_y],
\label{eq:sp_bc}
\eeq 
where $k_i = n_i \Delta k_i\, (i=x,z)$ are wavenumbers,  $n_i$ are integers, and
$\Delta k_i = 2\pi/L_i$. This condition is implemented by a few complex off-diagonal elements in the
compact-finite-difference matrices (see Ref.~\onlinecite{Schumann1985} and appendix B in \cite{Supplemental1}).

Because of the streamwise periodicity of the domain, the shift $S L_y t$ of the upper
boundary with respect to the bottom induces a characteristic time period, $ST_s= L_x/L_y$
(hereafter, box period), in which the flow is forced by the passing of the shifted-periodic
boxes immediately above and below.~\cite{Jimenez2013howlinear} Since this forcing acts at
the box scale, small turbulent structures can be expected to be roughly independent of the
box geometry, but care should be taken to account for resonances between the largest flow
scales and $T_s$ (see Sec.~\ref{sec4:OrrBursts}). Note that Eq.~(\ref{eq:bc1}) implies that
the flow becomes strictly periodic in $y$ whenever $t/T_s$ is integer. From now on, we will
use those moments as origins of time, and refer to them as the `top' of the box cycle.

Typically, codes that depend on shearing grids remesh them once per box period.~\cite{Rogallo1981}
The differences between those codes and the present one are
sketched in Fig.~\ref{fig:figgrid}. Consider the two-dimensional finite-differences
grid in Fig.~\ref{fig:figgrid}(a), which is drawn at some negative time, before
the top of the box cycle. The solid rectangle is the fundamental simulation box,
and the boundary condition \r{eq:bc1} is that the endpoints of the inclined solid
line have the same value. Without losing generality, this can be expressed by
writing the solution as $g=\widetilde{g}(x-Sty, y)$, where the dependence of
$\widetilde{g}$ on its second argument has period $L_y$. The first argument
represents the time-dependent tilting of the solution. Consider the representation of
$\widetilde{g}$ in terms of Fourier modes,
\beq
\widetilde{g}\sim \exp[\ii k_x(x-Sty) + \ii \widetilde{k}_y y] = 
   \exp[\ii k_x x + \ii (\widetilde{k}_y-Stk_x) y],
\label{eq:sp_bctilde}
\eeq
where periodicity requires that $\widetilde{k}_y = n_y \Delta k_y$. The effective
Fourier grid, $(k_x, k_y)=(n_x\Delta k_x, $ $n_y\Delta k_y- Stn_x\Delta k_x)$ is
skewed,~\cite{Rogallo1981} as represented by the upwards-sloping dashed lines in
Fig.~\ref{fig:figgrid}(b). Finite-differences formulas always use the Fourier modes
within the thick dashed rectangle (ABCD, plus its complex conjugate) in
Fig.~\ref{fig:figgrid}(b), whose limits are defined by the grid
resolution.~\cite{moin01} As the Fourier grid is sheared downwards for increasing
time, some modes leave the lower boundary of the resolution limit and are substituted
by others entering through the upper one. The overall resolution is therefore
maintained. In contrast, shearing-grid codes~\cite{Rogallo1981} let the Fourier modes
skew outside the resolution box between the periodic remeshings. The triangular
region (DcC) in Fig. \ref{fig:figgrid}(b) is thus not used, and the enstrophy
contained in it is either lost or aliased. In contrast, the triangular region (AbB)
is over-resolved, because it contains little useful information if the resolution has
been chosen correctly. Note that the Fourier modes just described do not correspond
to a unique time. It is easy to show that $t$ and $t+T_s$ result in exactly the same
Fourier grid, and are numerically indistinguishable. They typically correspond to
forward- and backwards-tilted pairs of boundary conditions, as shown by the the
dashed inclined line in Fig.~\ref{fig:figgrid}(a) and by the downwards-sloping single
Fourier grid line in Fig.~\ref{fig:figgrid}(b). In particular, it will become
important in \S\ref{sec:regen} that $t=\pm T_s/2$ form one such pair of
numerically-conjugate times.

Validations collected in Appendix C of the supplementary
material~\cite{Supplemental1} confirm that it maintains sixth-order accuracy in space and
third-order in time. It is also confirmed that it produces results that are consistent
with the well-known Rogallo's~\cite{Rogallo1981} three-dimensional spectral remeshing
method, both in the initial shearing of isotropic turbulence,~\citep{AGARD} and in longer
simulations of SS-HST.~\cite{Pumir1996,Gualtieri2002,Gualtieri2007} In all those cases,
our enstrophy and other small-scale statistics are slightly higher than those in remeshing
codes using the same Fourier modes, probably because the lack of remeshing prevents the
loss of some enstrophy.

\section{Characterization of SS-HST}\la{sec:boxes}

We have already mentioned that HST in an infinite domain evolves towards ever larger
length scales,~\cite{TavoularisKarnik1989,RogersMoin1985} so that simulations in a
finite box are necessarily controlled to some degree by the box geometry. In this
section, we examine the dependence of the turbulence statistics on the geometry and
on the Reynolds number. Our aim is to identify a range of parameters in which
the flow is as free as possible from box effects and can be used as a model
of shear-driven turbulence in general.
 
A wide range of aspect ratios was sampled at Reynolds numbers between $Re_z=1000$ and
48000. Each simulation accumulates statistics over more than $St=100$, after
discarding an initial transient of $St\approx 30$. The lowest Reynolds number at
which statistically steady turbulence can be achieved is $Re_z \approx 500$, but we
do not consider the transitional low-Reynolds number regime, and focus on the
fully-developed cases in which turbulence does not laminarize for at least $St>100$.
Most of our simulations have well-resolved numerical grids with effective resolutions
$\Delta x_i \lesssim 2 \eta$ in the three coordinate directions. A few very flat
boxes with $A_{yz} < 0.5$ have coarser grids with $\Delta x_i \approx 6 \eta$, since
it will be seen that those geometries are severely constrained by the box, and are
not interesting for the purpose of this paper. One of those cases was repeated at
$\Delta x_i \approx 2 \eta$ to confirm that the one-point statistics vary very little
for the coarser grids. Also, because of the computational cost, some long- and
tall-box cases, $A_{xz} = 8$, $A_{yz} \gtrsim 1 $ and $Re_z > 3000$, were run at
$\Delta x_i \approx 3\eta$, which is empirically the marginal value to investigate
small scales.~\cite{jwsr} In all cases, the energy input $\mathcal{P}$ balances the
dissipation rate $\diss$ within less than $5\%$.
 
\begin{figure}[t]
\centering
  \includegraphics[width=0.90\linewidth,clip]{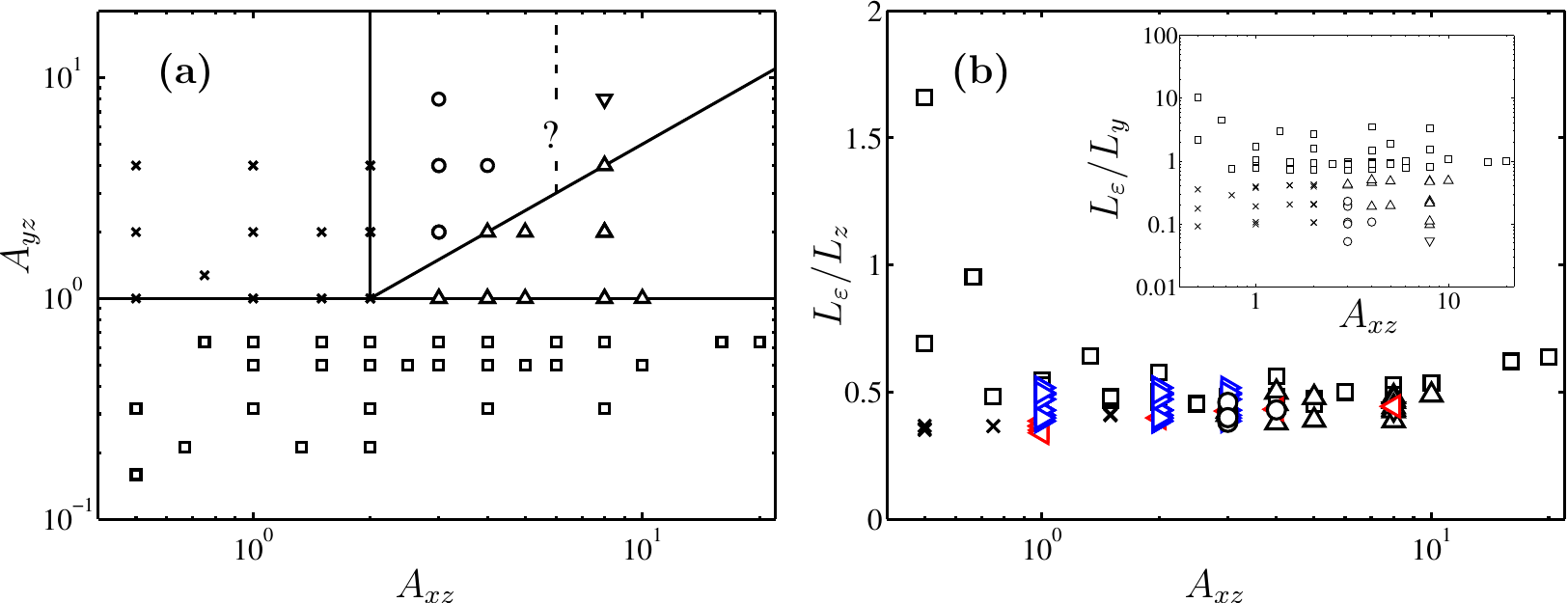}%
  \caption{%
(a) The aspect ratios, $(A_{xz}, A_{yz})$, of the DNSes used in the paper, and the
definition of the classification of the computational boxes into different regimes.
(b) Integral scale $L_{\varepsilon}/L_z$ as a function of $A_{xz}$ in
semi-logarithmic scale. The inset shows $L_{\varepsilon}/L_y$ in logarithmic scale.
Symbols as in Table~\ref{tab:zones}, but $\vartriangleleft$ from Ref.~\onlinecite{Pumir1996}
and $\vartriangleright$ from Refs.~\onlinecite{Gualtieri2002,Gualtieri2007}.
}
\label{fig:symbols} 
\end{figure}

It turns out that the effect of the Reynolds number is relatively minor, at least for
the larger scales, but that the flow properties depend strongly on the aspect ratios.
A summary of the geometries tested is Fig.~\ref{fig:symbols}(a). The plane is divided
into several regions, whose properties will be shown to be different. Those ranges,
as well as the associated symbols used in the figures and the short name by which we
refer to them below, are summarized in Table~\ref{tab:zones}. The characteristics of
a few representative simulations can also be found in Table~\ref{table:HST_Axz3} in
\S\ref{sec:ssp}.

\begin{table}[t]
\caption{Parameter ranges in the space of aspect ratios for the SS-HST simulations}%
\vspace{0.5ex}\la{tab:zones}
\centering
\begin{tabular}{lccccc}
\hline\hline
Region & Symbol &   &  \\
 \hline
Short       &  $\mathbf{\times}$   & $A_{xz} \le 2$      &  $A_{yz} \ge 1$ \\
Flat          &  $\square$          & $-$  &  $A_{yz} < 1$ \\
Acceptable &  $\circ$               & $2 < A_{xz} $  ($\lesssim 5$) &  $ A_{xz} < 2A_{yz}$ \\
Long        &  $\vartriangle$     & $A_{xz} \ge 2A_{yz}$     &  $A_{yz} \ge 1$ \\
Tall and long &  $\triangledown$     & $A_{xz} (\gtrsim 5)$ & $A_{xz} < 2A_{yz}$  \\
\hline\hline
\end{tabular}
\end{table}

\begin{figure}[tbp]
  \centering
   \includegraphics[width=0.9\linewidth,clip]{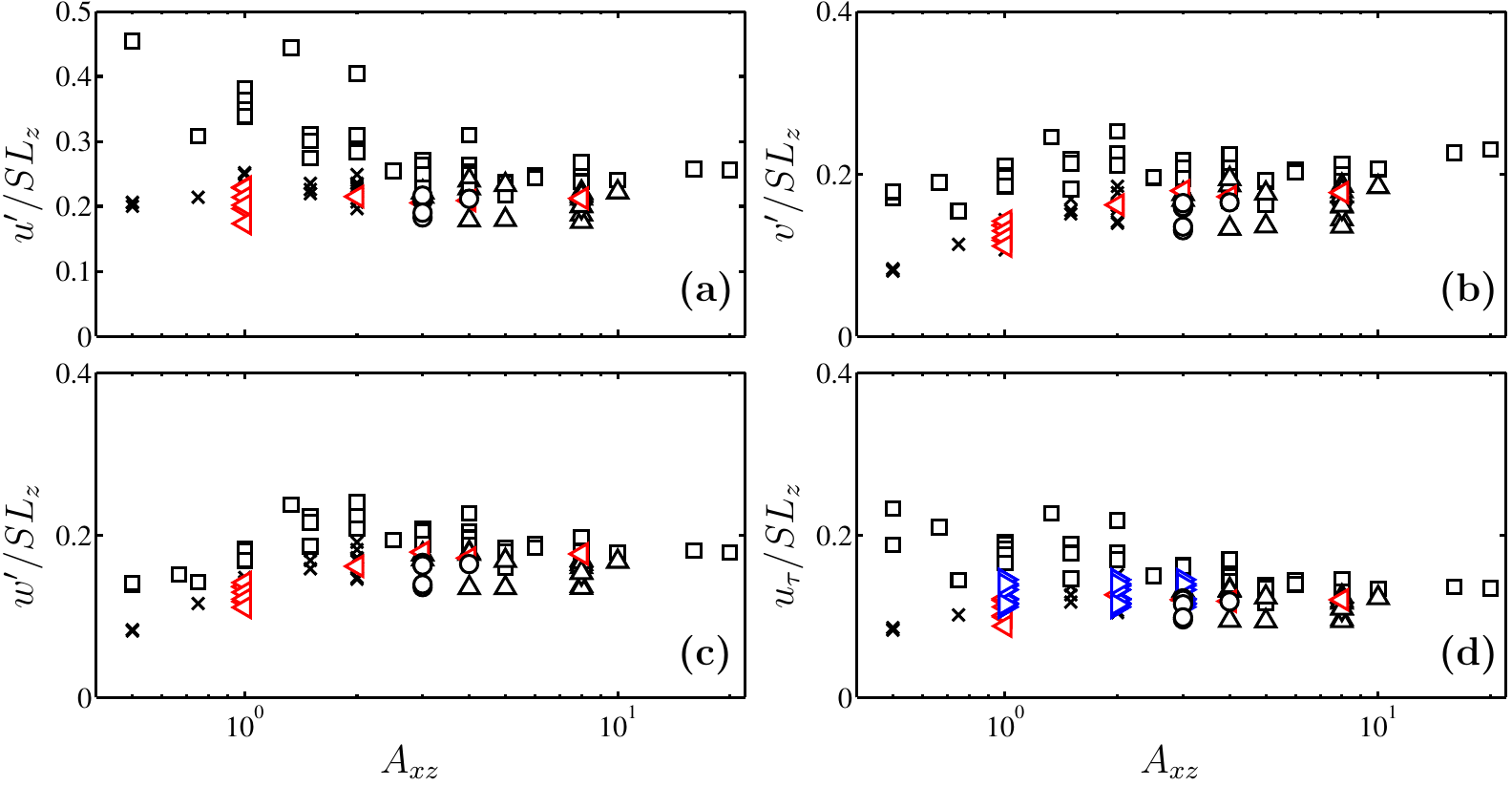}
\caption{Velocity fluctuation intensities as functions of $A_{xz}$, scaled with $SL_z$: 
(a) $\urms/SL_z$. (b) $\vrms/SL_z$. (c) $\wrms/SL_z$. 
(d) Tangential Reynolds stress, given as a friction velocity.
Symbols as in Table~\ref{tab:zones}, but $\vartriangleleft$ from Ref.~\onlinecite{Pumir1996} and $\vartriangleright$ 
from Refs.~\onlinecite{Gualtieri2002,Gualtieri2007}.
}\label{fig:uvw}
\vspace{2mm}
  \centering
   \includegraphics[width=0.9\linewidth,clip]{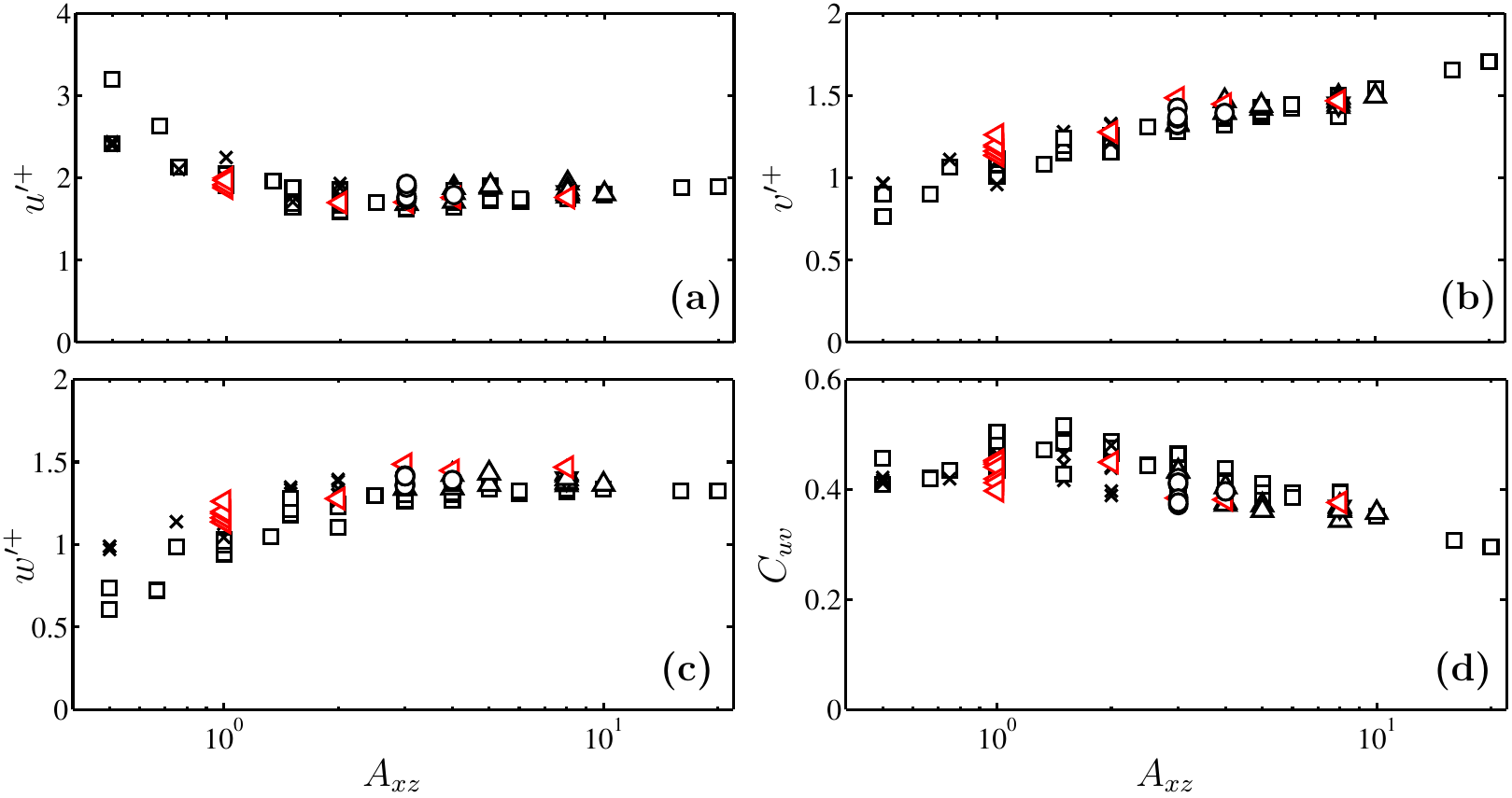}
  \caption{Velocity fluctuation intensities as functions of $A_{xz}$, scaled with the friction velocity. 
(a) $\upls$. (b) $\vpls$. (c) $\wpls$.
(d) Structure coefficient $C_{uv} = -\bra u v\ket/\urms\vrms$.
Symbols as in Table~\ref{tab:zones}, but $\vartriangleleft$ from Ref.~\onlinecite{Pumir1996}.
}\label{fig:uvw_plust}
\end{figure} 

\subsection{The characteristic length and velocity}\label{sec4:onepoint}

Although it is reasonably well understood that the time scale of sheared turbulence
is $S^{-1}$, there is surprisingly little information on the length scales. Sheared
turbulence itself has no length scale, except for the viscous Kolmogorov length, and
properties like the integral scale grow roughly linearly in
time.~\cite{RogersMoin1985,TavoularisKarnik1989} In statistically stationary
simulations, the question is which of the box dimensions limits the growth and
imposes the typical size of the turbulent structures. Fig.~\ref{fig:symbols}(b) shows
the integral scale, $L_{\varepsilon} \equiv (q^2/3)^{3/2}/\diss$, normalized by $L_z$ as a
function of $A_{xz}$. The inset in that figure tests the scaling of $L_{\varepsilon}$
with $L_y$, and it is clear that $L_z$ is a much better choice. The ordinates in the
inset are logarithmic, and $L_{\varepsilon}/L_y$ spans two orders of magnitudes,
while $L_{\varepsilon}/L_z$ is always around 0.5, except for a few short-flat cases.
This suggests that the flow is `minimal' in the spanwise direction, and visual
inspection confirms that there is typically a single streamwise streak of $u$ that
fills the domain (see Fig.~\ref{fig:q2q4} in Sec.~\ref{sec:ssp}). A similar structure
is found in minimal channels, both near the wall \cite{JimenezMoin1991} and in the
logarithmic layer,~\cite{FloresJimenez2010} and Pumir~\cite{Pumir1996} reports that
most of the kinetic energy in his SS-HST is contained in the first spanwise
wavenumber. The scaling with $L_x$, not shown, is even worse than with $L_y$.

The scaling of the length suggest that velocities should scale with $S L_z$. 
This is tested in Fig.~\ref{fig:uvw}(a-c) for the velocity fluctuation intensities. 
They collapse reasonably well except for very flat boxes, $A_{yz}<1$. The scaling
with $SL_y$, not shown, is very poor.

The tangential Reynolds stress is given in Fig. \ref{fig:uvw}(d), in the form of a
friction velocity defined as $u_\tau^2 = - \bra u v \ket + \nu S$. As was the case
for the intensities, it scales well with $SL_z$, but it turns out that most of the
residual scatter in Fig. \ref{fig:uvw} is removed using $u_\tau$ as a velocity
scale, as shown in Fig. \ref{fig:uvw_plust}(a-c). It is interesting that, when the
intensities are scaled in this way, they are reasonably similar to those in the
logarithmic layer of wall-bounded flows, which are $(u'^+,v'^+,w'^+)\approx
(2,1.1,1.3)$ in channels.~\cite{HoyasJimenez2006} The usual argument to support this
`wall-scaling' is that the velocity fluctuations have the magnitude required to
generate the tangential Reynolds stress, $u_\tau^2$, which is fixed by the momentum
balance. The argument also requires a constant value of the anisotropy of the
Reynolds-stress tensor, whose most important contribution is the structure coefficient
$C_{uv}=-\bra u v \ket/ \urms \vrms$. It is given in Fig.~\ref{fig:uvw_plust}(d), and
also collapses well among the different cases, although it varies slowly with
$A_{xz}$. Note the good agreement of the intensities in Figs.~\ref{fig:uvw} and
\ref{fig:uvw_plust} with the results included from previously published simulations.~\cite{Pumir1996,Gualtieri2002,Gualtieri2007}
 
We defer to the next section the discussion of the slow growing trend of $v'^+$ with the
aspect ratio, but the behavior of the three intensities for short boxes is interesting and
probably has a different origin. For $A_{xz}\lesssim 1$, $u^+$ ($v^+$, $w^+$) is stronger
(weaker) than the cases with more equilateral boxes. A similar tendency was observed in
simulations of turbulent channels in very short boxes by Toh and Itano.~\cite{TohItano2005}
It can also be shown that the inclination angle of the two-point correlation function of $u$
tends to zero for these very short boxes, while it is around $10^\circ$ in longer boxes (not
shown), and in wall-bounded flows.~\cite{SilleroJimenezMoser2014} The behavior of the
intensities suggests that the streaks of the streamwise velocity become more stable and
break less often when the box is short, which is reasonable if we assume that the box limits
the range of streamwise wavenumbers available to the instability. Indeed, the linear
stability analysis of model streaky flows shows that they are predominantly unstable to long
wavelengths, and that shorter instabilities require stronger streaks.~\cite{Waleffe2003}

%
\begin{figure}[tb]
\centering  
\includegraphics[width=0.9\linewidth,clip]{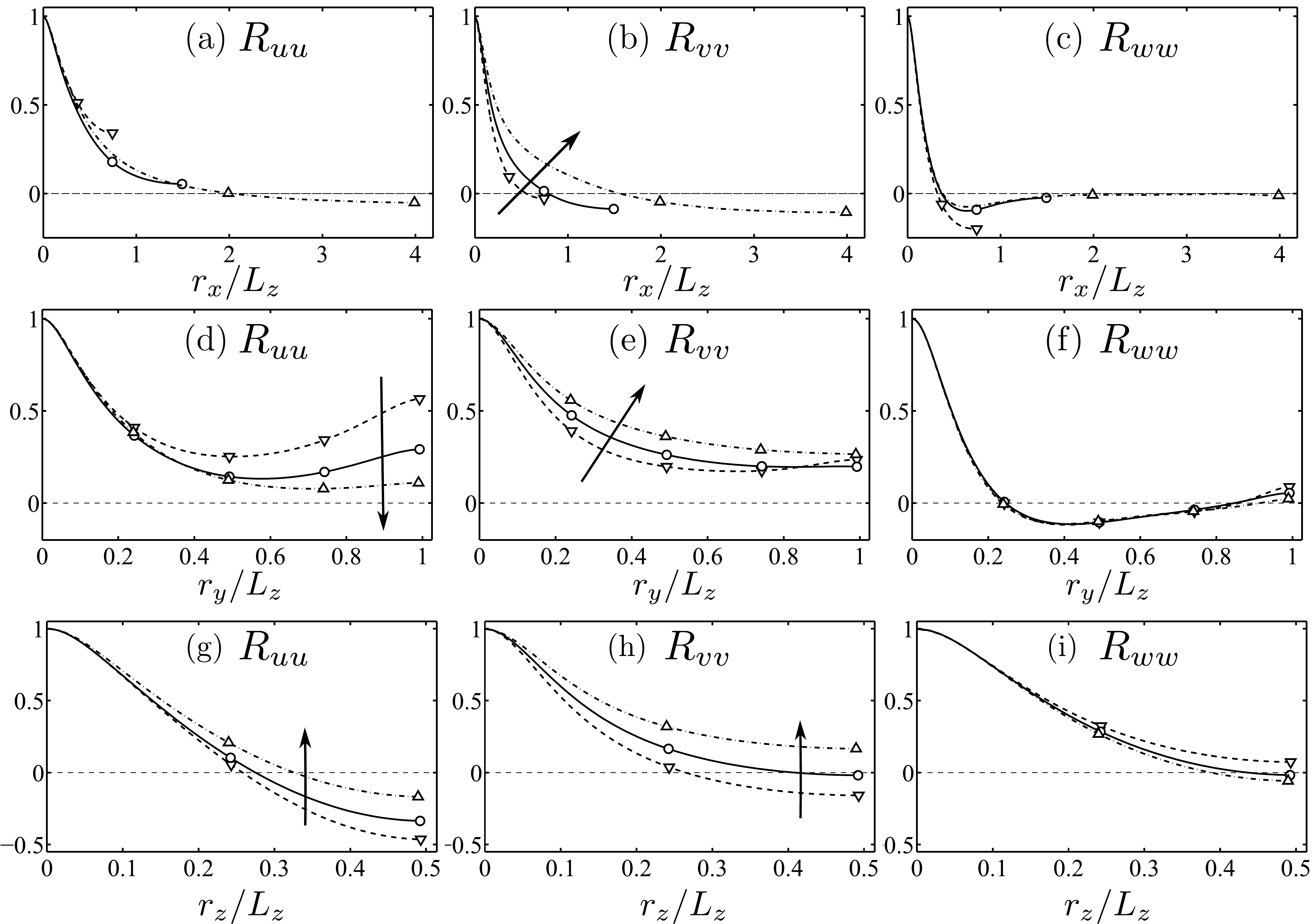}%
\caption{%
Two-point velocity correlation functions: (a,d,g) $R_{uu}$; (b,e,h) $R_{vv}$;
(c,f,i) $R_{ww}$, in (a-c) streamwise, (d-f) vertical, and (g-i) spanwise directions, for
$Re_z=2000$ and $A_{yz}=1$.
\dashedtridown, $A_{xz}=1.5$; \solidcirc, 3; \chndottri, $8$.
The arrows are in the sense of increasing  $A_{xz}$.
   }
  \label{fig:corline1-1} 
\vspace{2mm}%
  \centering  
  \includegraphics[width=0.9\linewidth,clip]{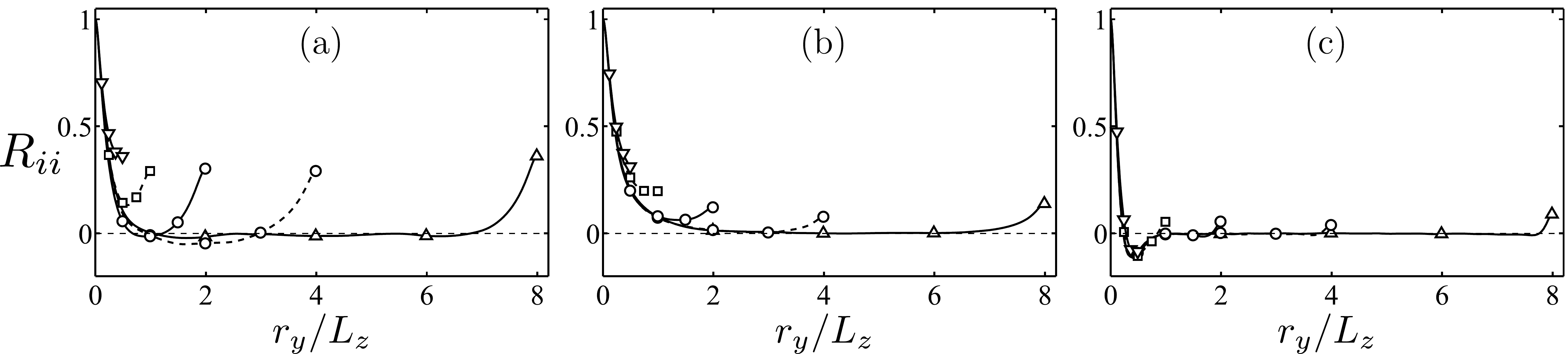}\hspace*{2mm}
  \caption{%
  Two-point velocity correlation functions: (a) $R_{uu}$, (b) $R_{vv}$, and (c) $R_{ww}$, for
   $Re_z=2000$ and $A_{xz} = 3$.
  \solidtridown, $A_{yz}=0.5$; \dashedsquare, 1;  \solidcirc,  2; \dashedcirc, 4; \solidtri, 8.
  }
  \label{fig:corline2-4} 
\end{figure}

Figs.~\ref{fig:corline1-1} and \ref{fig:corline2-4} display the two-point
correlation functions of the velocities. The streamwise correlations in Fig.
\ref{fig:corline1-1}(a-c) are the longest, especially $R_{uu}(r_x)$, 
as already noted by Ref.~\onlinecite{RogersMoin1985}. 
This is common in other shear flows,~\cite{SilleroJimenezMoser2014} 
and Fig.~\ref{fig:corline1-1}(a) shows that at least
$A_{xz}=2$ is required for the streamwise velocity to become independent of the box
length. This roughly agrees with the observation of Reynolds-stress structures in
channels, which have aspect ratios of the order of $L_x/L_y\approx 3$ (see Ref.~\onlinecite{LozanoFloresJimenez2012}), 
and with our discussion on the velocity intensities
in short boxes. Pumir~\cite{Pumir1996} likewise reported that the turbulence
statistics become independent of the aspect ratio for $A_{xz} \gtrsim 3$, and Rogers
{\it et al.}~\cite{RogersMoin1985} used $A_{xz}=2$ for their transient-shearing
experiments after a qualitative inspection of the correlation functions. It is
interesting that $R_{vv}$ in Figs. \ref{fig:corline1-1}(b,e,h) becomes longer in the
three directions for longer boxes, even if $v$ is typically a short variable in
wall-bounded flows.~\cite{Jimenez2013nearwall,SilleroJimenezMoser2014} 
This agrees with the common notion that the limit on the size of $v$ in wall-bounded turbulence
is due to the blocking effect of the wall, and suggests that the effect observed in
the longer boxes is related to structures that span several vertical copies of
the shear-periodic flow. Otherwise, the central peak of all the correlations
collapses well when normalized with $L_z$, even if $A_{xz}$ spans a factor of over 20
in our simulations. Note that none of the curves in Fig.~\ref{fig:corline1-1}(g,h,i) 
shows signs of decorrelating at the box width, supporting the
conclusion that the flow is constrained by the spanwise box dimension.
 
Fig.~\ref{fig:corline2-4} shows two-point velocity correlation functions in the
$y$ direction, for different vertical aspect ratios. The collapse with $L_z$ of the
different correlations is striking; taller boxes lead to more structures
of a given size, but not to taller structures. The figure also provides a minimum
vertical aspect ratio, $A_{yz}\approx 2$, for structures that are not vertically
constrained, ruling out what we have termed above `flat' boxes.
 
The correlation functions in Fig.~\ref{fig:corline1-1} and \ref{fig:corline2-4}
depend on the Reynolds number in a natural way. The inner core narrows as the
Reynolds number increases and the Taylor microscale decreases, but the outer part of
the correlations stays relatively unaffected. 

The Taylor-microscale Reynolds number satisfies $Re_\lambda \approx Re_z^{1/2}$
except for very flat or short boxes. This agrees approximately with Pumir's
data~\cite{Pumir1996} and the latest high-Reynolds-number DNS by Gualtieri, {\it et
al.}~\cite{Gualtieri2007}, although their earlier data~\cite{Gualtieri2002} are
10--20\% smaller than in our simulations. The difference is too small to decide
whether the reason is numerical or statistical.

\subsection{Long boxes, the bursting period, and box resonances}\label{sec4:OrrBursts}
%
\begin{figure}[t]
\centering
\includegraphics[width=0.95\linewidth]{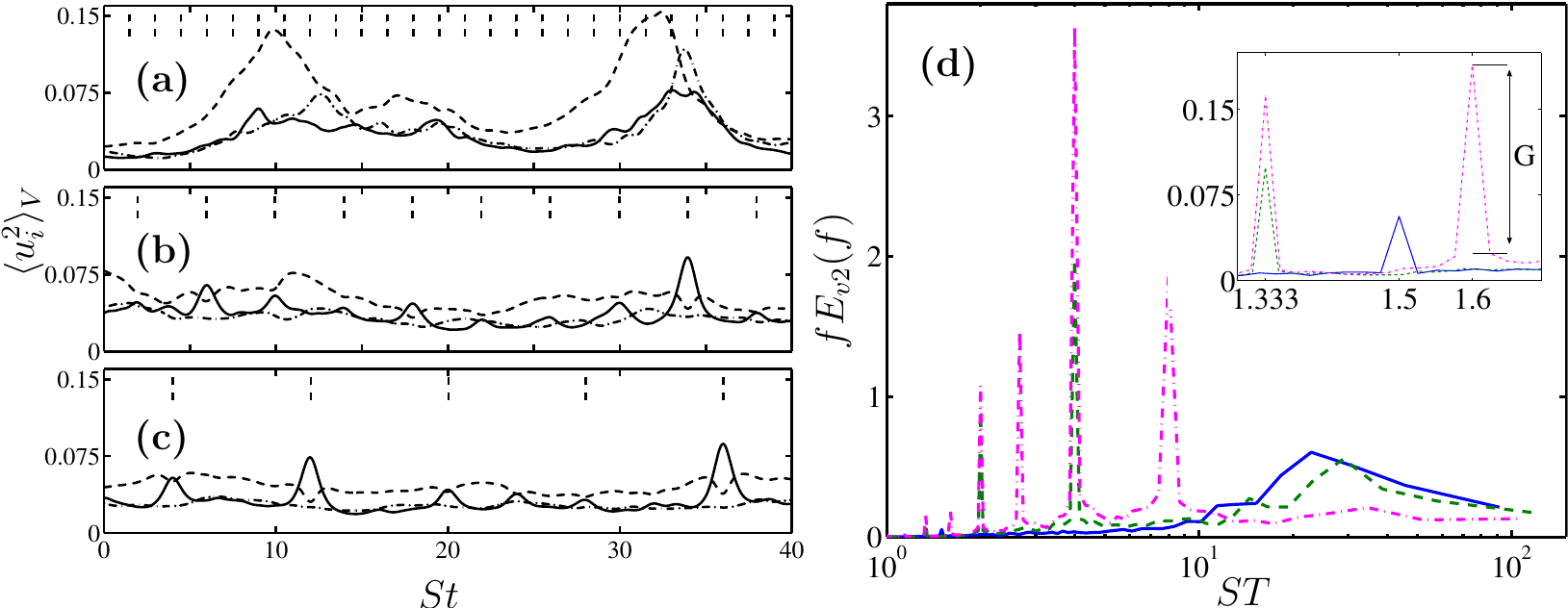}
\caption{
(a-c) The time history of box-averaged squared velocity intensities, normalised by
$(SL_z)^2$: \dashed, $u^2(t)$; \solid, $v^2(t)$; \chndot, $w^2(t)$. 
(a) $A_{xz} = 1.5$; (b) 4; (c) 8. In all cases, $Re_z=2000$, $A_{yz}=1$. The vertical
dashed lines mark the top of the box cycle.
(d) Premultiplied frequency spectra of $v^2(t)$ 
in (a-c), $f E_{v2}(f)$, as functions of the period $T =2\pi/ f$. \solid, (a); \dashed,
(b); \chndot, (c). The inset is a zoom of the region $1.3 \le ST \le 1.7$, and defines the amplification
in Fig. \ref{fig:pre_fre_tb_g}(c).
} 
\label{fig:traces}
\end{figure}
%
The results up to now show that statistically-stationary shear flow is
mostly limited by the spanwise dimension of the box, but that very short,
$A_{xz}\lesssim 2$, or very flat boxes, $A_{yz}\lesssim 1$, should be avoided
because they are also minimal in the streamwise or vertical direction. 
Figs.~\ref{fig:uvw_plust} and \ref{fig:corline1-1} show that there is an additional effect
of very long boxes, which have stronger and longer fluctuations of $v$ and, to
some extent, of $u$. This is the subject of this section.

The most obvious property of the SS-HST flows is that they burst intermittently. This
is seen in the time histories of the box-averaged velocity fluctuations, given in
Fig.~\ref{fig:traces}(a-c) for three boxes of different streamwise elongation.
The character of the bursting changes with $A_{xz}$. For the relatively short box in
Fig.~\ref{fig:traces}(a) the bursts are long, and predominantly of the streamwise
velocity $\bra u^2\ket_V$. They are followed some time later by a somewhat weaker
rise of $\bra v^2\ket_V$ and $\bra w^2\ket_V$. This relation between 
velocity components is reminiscent of the bursting in minimal channels analyzed 
in Ref.~\onlinecite{Jimenez2013howlinear}, and the bursting time scale ($ST\approx 15$--20) is also
of the same order. The longer boxes in Fig.~\ref{fig:traces}(b,c) undergo a
different kind of burst, sharper $(ST\approx 2)$, and predominantly of $\bra
v^2\ket_V$. These sharp bursts always occur at the top of the box cycle, which is
marked by the vertical dashed ticks in the upper part of the figures. There is little
sign of interaction of the box cycle with the history in Fig.~\ref{fig:traces}(a).

The frequency spectra of the histories of $\bra v^2\ket_V$ are given in Fig.
\ref{fig:traces}(d). They have two well-differentiated components: a broad peak
around $ST=25$ that corresponds to the width of the bursts in Fig.~\ref{fig:traces}(a), and
sharp spectral lines that represent the average distance between the shorter bursts in  Fig.~\ref{fig:traces}(b,c). We will see below that these lines are at the box period and its
harmonics. Note that these are not frequency spectra of the velocity $v$, but of
the temporal fluctuations of its box-averaged energy, and that their dimensions are 
proportional to $v^4$. 
Note also that the broad `turbulent' frequency peak is still present in the resonant
cases, although it weakens progressively as the resonant component takes over.

\begin{figure}[t]
\centering
\includegraphics[width=0.95\linewidth]{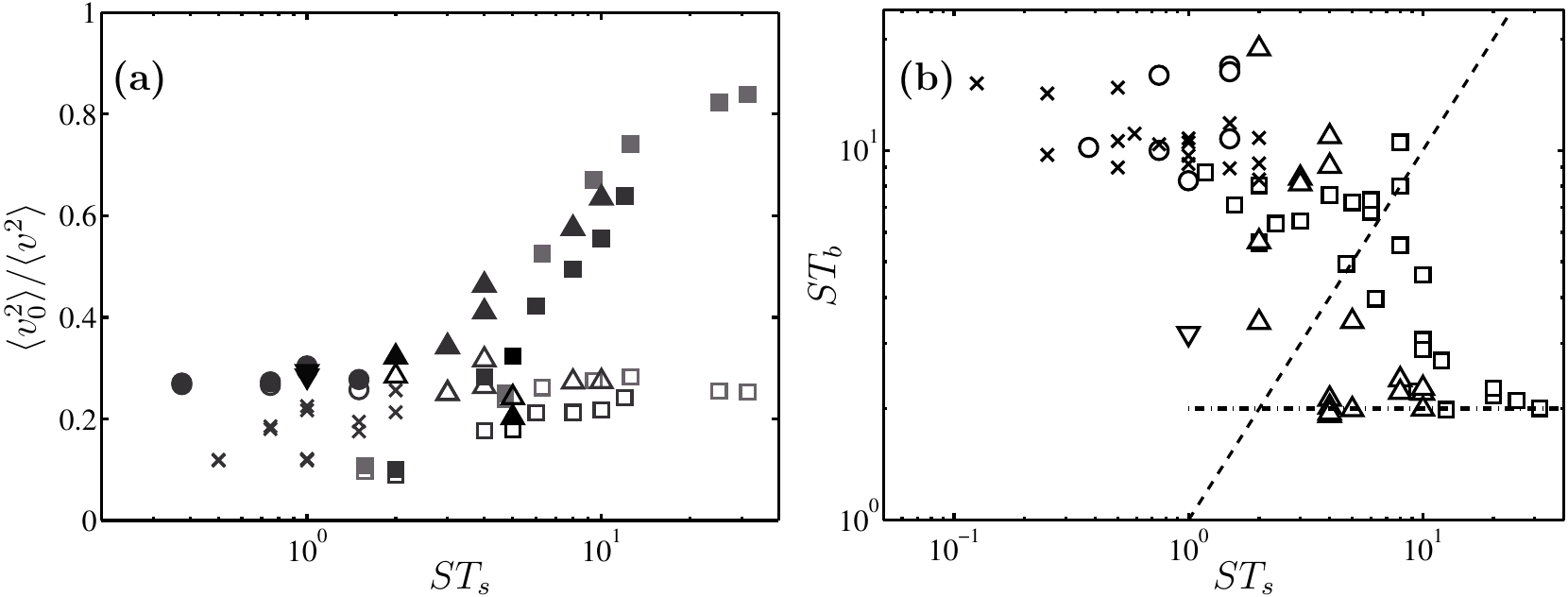}
\caption{
(a) Energy in the first six two-dimensional modes $(k_z=0)$ of $v$ as a function of
the box period $ST_s = A_{xz}/A_{yz}$. Several values of $A_{yz}$ are included. Open
symbols are averaged over the whole history. Solid ones are conditionally averaged at
the top of the box cycle. Symbols are gray-colored by the Reynolds number, from
$Re_z=1000$ (light-gray) to 3200 (dark-gray).
(b) Bursting time scale, $ST_b$, obtained from the width of the temporal autocorrelation
function of $\bra v^2\ket_V$. The diagonal dashed line is $T_b=T_s$; the horizontal
one is the bursting width of a single linearized Orr burst.
Symbols as in Table~\ref{tab:zones}.
} 
\label{fig:ene_10}
\end{figure}

The resonant lines correspond to the narrow bursts of Fig.~\ref{fig:traces}(b,c),
and get stronger with increasing $A_{xz}$. Fig.~\ref{fig:ene_10}(a) shows that they
are characterized by a temporary increase in the two-dimensionality of the flow. The
open symbols in that figure show the fraction of $\bra v^2\ket$ contained in the
first few two-dimensional $x$-harmonics $(k_z=0, k_x=2\pi n_x/L_x, n_x=1,\ldots, 6)$,
plotted against $ST_s=A_{xz}/A_{yz}$. They account for a relatively constant fraction
(approximately 20\%) of the total energy. The solid symbols are the same quantity
computed at the top of the box cycle, $t=n T_s$. It begins to grow at elongations of the
order of $ST_s\approx 2$, and accounts for almost 60\% of the total energy for the
longest boxes. Note that the data collapse relatively well with $ST_s$, rather than
with $A_{xz}$.

The growth of the two-dimensionality interferes with the overall behavior of the flow. The
average length $T_b$ of individual bursts is given in Fig. \ref{fig:ene_10}(b),
computed as the width of the temporal autocorrelation function of $\bra v^2\ket_V$, measured
at $C_{v^2v^2}=0.5$ (see Ref.~\onlinecite{Jimenez2013howlinear}). For short boxes
$(ST_s\lesssim 2)$, the period stays relatively constant, $S T_b \approx 15$. But it
decreases when the two-dimensionality begins to grow in Fig. \ref{fig:ene_10}(a). The
diagonal dashed line in Fig.~\ref{fig:ene_10}(b) is $T_b=T_s$, and strongly suggests that
the effect of the long boxes is associated with the interaction of the box period with the
bursting. Note that the bursting width for $ST_s\gg 2$ tends to be that of a linearized
two-dimensional Orr burst.~\cite{Orr1907}

\begin{figure}[t]
\centering
\includegraphics[width=0.95\linewidth]{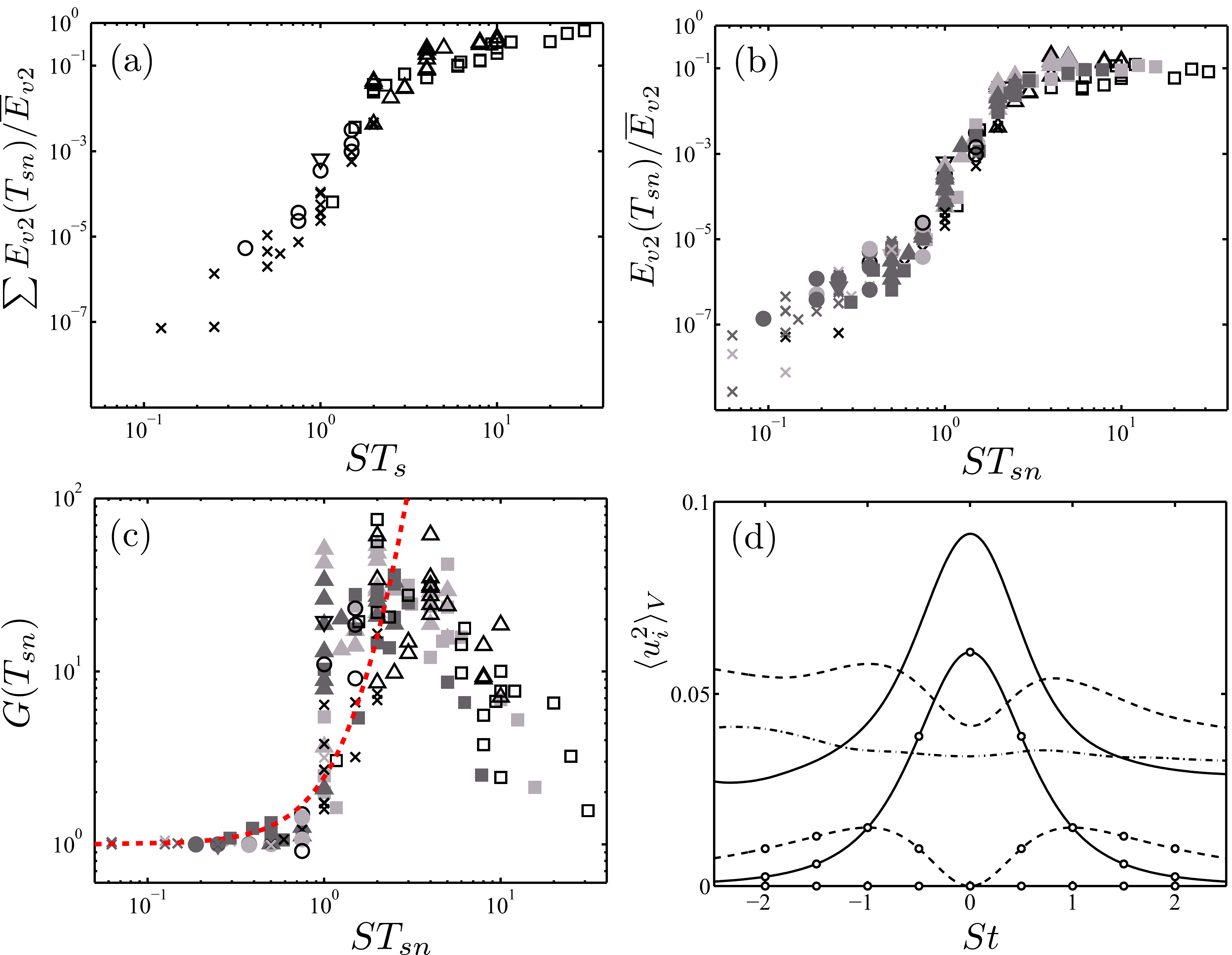}
\caption{
(a) Total spectral energy in the harmonics of the box period in the frequency
spectrum of $\bra v^2\ket_V$, normalized with the total energy, and versus the box period. 
Symbols as in Table~\ref{tab:zones}.
(b) Spectral energy in each of the harmonics of the box period, $T_{sn}= T_s/n_x$ ($n_x=1,2$ and 4), 
as in (a). Open symbols are for $n_x=1$; grey, $n_x=2$; black, $n_x=4$.
(c) The amplification $G$ of the individual harmonics of the box period, as defined
in the inset in Fig.~\ref{fig:traces}(d). Symbols as in (b). The dashed line is Eq.~(\ref{eq:G}).
(d) Lines without symbols are the segment of the history of integrated intensities in
Fig.~\ref{fig:traces}(b), around $St= 35$. For comparison, the lines with
symbols are the inviscid linearized solution (see Eq.~\ref{eq:rdt_sol_1}) of the fundamental mode
$k_x=2\pi/L_x, k_z=0$, scaled to the same amplitude of the peak of 
the corresponding mode, $\widehat{v}_{10}$, of DNS.
\dashed, $\bra u^2\ket_V$; \solid, $\bra v^2\ket_V$; \chndot, $\bra w^2\ket_V$.
} 
\label{fig:pre_fre_tb_g}
\end{figure}

That the spikes are associated with the box period is shown 
in Fig.~\ref{fig:pre_fre_tb_g}(a), which displays the total spectral energy of the temporal evolution
of $\bra v^2\ket_V$ at period $T=T_s$ and its harmonics $T_{sn}=T_s/n_x$. It is
negligible for boxes with $ST_s\lesssim 1$, but increases rapidly after that threshold.
It then keeps increasing slowly until most of the $\bra v^2\ket$ fluctuations are due to the
spectral spikes.

The reason for the residual growth after resonance is seen in
Fig.~\ref{fig:pre_fre_tb_g}(b), where the energy in each harmonics is plotted separately
versus $T_{sn}$. Each harmonic resonates with its own box period, $ST_{sn} \approx 1$,
after which its amplitude decreases slightly (note the wide vertical scale of these two
figures). The slow growth of the spike energy beyond the resonance in
Fig.~\ref{fig:pre_fre_tb_g}(a) is due to the accumulation of new resonant harmonics.

In fact, if we accept that the spikes are due to the amplification of pre-existing
background fluctuations, Fig.~\ref{fig:pre_fre_tb_g}(c) shows that the maximum
amplification takes place at $ST_{sn} \approx 1$. This is true for each individual
harmonic. The quantity $G \equiv E_{v2} (T_{sn})/ \widetilde{E}_{v2} (T_{sn})$
in this figure is sketched in the inset of Fig.~\ref{fig:traces}(b), and is the ratio
between the energy in the spectral spike, and the energy in the background spectrum at the
same frequency when the spike is removed by linear interpolation across the
sharp spectral line, $\widetilde{E}_{v2} (T_{sn}) = (E_{v2} (T_{sn} + \Delta T) +
E_{v2} (T_{sn} - \Delta T) )/2$.

It is tempting to hypothesize that a phenomenon with the same periodicity as the box-passing
period is due to the interaction between structures from neighboring shear copies.
Fig.~\ref{fig:pre_fre_tb_g}(d) shows that this is not the case. It displays the details of
the velocity amplitudes during one of the spikes, and it agrees almost exactly with the
linearized Orr burst included in the figure as a comparison.~\cite{Orr1907} The overtaking
of shear-periodic copies advected by the mean velocities in neighboring boxes is not very
different from the shearing mechanism in Orr's. For example, the characteristics dip in
$u^2$ during the burst of $v^2$ is the same in both
cases,~\cite{UmurhanRegev2004,Jimenez2007,Jimenez2013howlinear} and the essence of the
interaction is the tilting of the structures due to the shear. However, the details are
different. In the case of the shear-periodic boundary condition, the interacting structures
pass each other at a constant vertical distance. It can then be shown that both the period
and the width of the velocity fluctuations are of order $T_s$, and that their amplitude
decays very fast as $L_y/L_x$ increases, essentially because the shear-periodic
copies move farther from each other.\cite{Jimenez2013howlinear} In the case of the shear,
the mechanism is local tilting, and the time scale of the amplification is $S^{-1}$.
The width of each burst is $St=O(1)$, but the distance between bursts is still
$T_s$. It is clear from Fig.~\ref{fig:pre_fre_tb_g}(d) that this is the case here.
Inspection of the sharp bursts in Figs. \ref{fig:pre_fre_tb_g}(b,c) shows that their width
is a few shear times, independently of $T_s$.

\begin{figure}[t]
\centering
\includegraphics[width=0.60\linewidth]{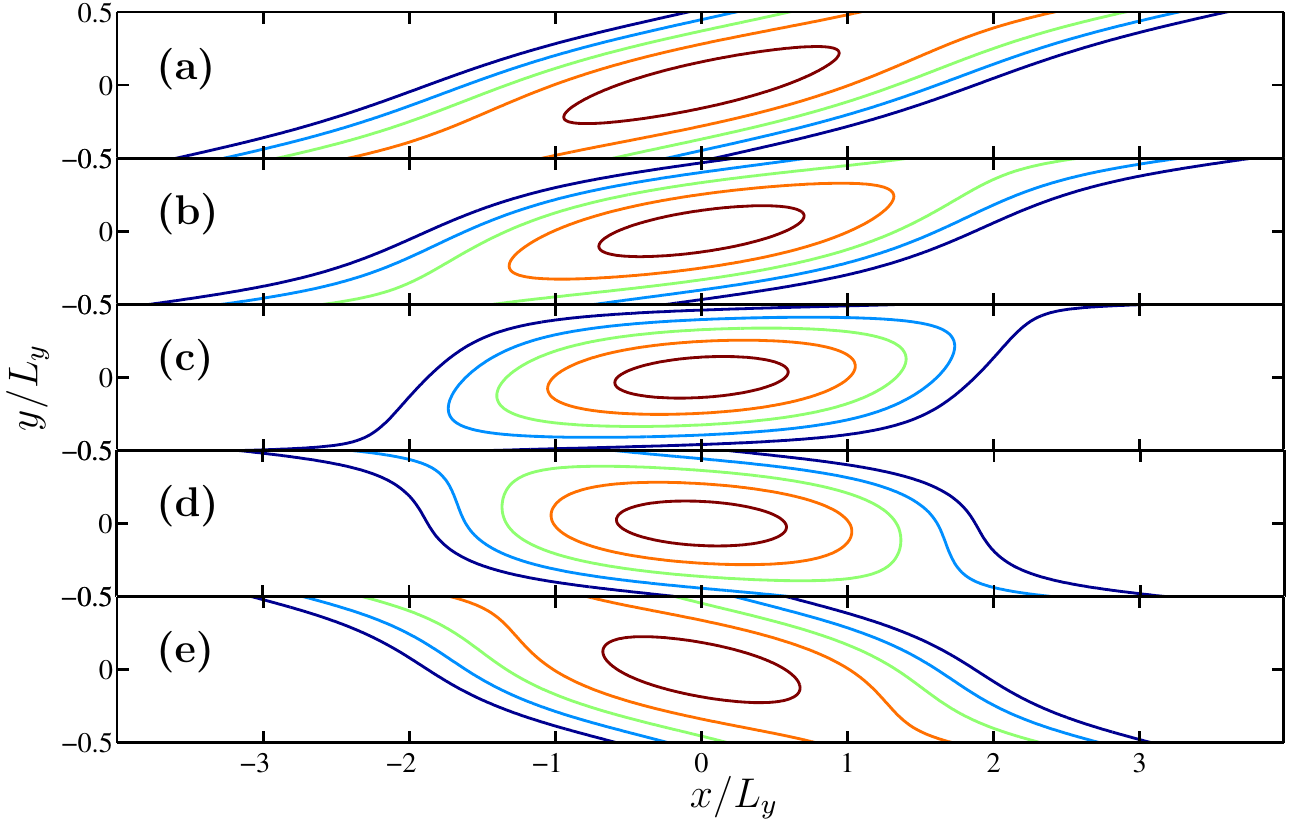}
\caption{
Two-point spatial correlation function of $\hv_{10}$.  Frame (c) is the bottom of the box
cycle, $t=T_s/2$. Time is from top to bottom, separated by $St=0.5$
between frames. Isolines are $R_{vv}=$(0.1:0.2:0.9). $Re_z=2000,\, A_{xz}=8,\,
A_{yz}=1$.
} 
\label{fig:movie2D}
\end{figure}

On the other hand, Orr bursts are not periodic. The Orr mechanism is a linear
amplification process that works on preexisting initial conditions of the right shape
(backwards tilting wavefronts), even when they are only one component of a more complex
flow. In turbulence, these conditions are created `randomly' by nonlinear processes, and
Orr-like bursts, not necessarily two-dimensional, occur whenever that happens. But the
observed two-dimensional quasiperiodic bursting with the passing frequency of the box
requires a more deterministic explanation. We have seen that the flow during the spikes is
two-dimensional in the $(x,y)$ plane. Taking the origin of time at the top of the burst, the
inviscid solution of the linearized equations (see Ref.~\onlinecite{Jimenez2013howlinear})
for two-dimensional perturbations with $k_z=0$ and $k_y(0)=0$ is a single
burst,\cite{Moffatt1967,Townsend1976book}
\begin{equation}\label{eq:rdt_sol_1}
  \widehat{v}(t)/\widehat{v}_0 = (k_x^2 + k_{0y}^2 + k_z^2)/(k_x^2 + k_y^2 + k_z^2) = (1+S^2 t^2)^{-1},
\end{equation} 
where $\widehat{v}_0$ is the initial vertical velocity at the top of the burst, 
$k_{0y}$ and $k_y = k_{0y} - Sk_x t$ are the initial and time-evolving wave numbers.
Fig.~\ref{fig:movie2D} shows the the evolution near the bottom of the box cycle $(t\approx
T_s/2)$ of the two-point correlation function of the $(n_x=1, n_z=0)$ harmonic $\hv_{10}$
within the fundamental numerical box. A band of $v$ is tilted forward, as in an infinite
shear, and can only extend across the box boundaries by connecting with one of the several
periodic copies of itself in the boxes immediately above and below the fundamental box. It
does so linking with the closest copy. Consider the top boundary. In the first part of the
box cycle, $t< T_s/2$, the upper-box structure closest to the fundamental one is the one
ahead of it, and the band keeps evolving as an infinitely long forward-tilted sheet. At
$t=T_s/2$ the distance from the fundamental to the two copies ahead and behind it is the
same, and after that moment the trailing copy is the closest one. The figure shows that the
correlation condenses into an oval structure and reconnects itself to the trailing copy. The
result is a backwards-tilting layer that restarts the shearing amplification cycle.

A rough estimate of the amplifications in Fig.~\ref{fig:pre_fre_tb_g}(c) can be obtained
from this mechanism. The amplitude of $v$ in an Orr burst, $v\propto \cos^2 \psi$, depends
almost exclusively of the front inclination angle $\psi = \arctan(k_y/k_x)$ (see
Ref.~\onlinecite{Jimenez2013howlinear}). The tilting of the velocity fronts at the moment of
reconnection is given by $\tan \psi_r=-L_x/2L_y=-ST_s/2$. The vertical velocity $v_r$ is
amplified as the layer moves towards the vertical ($\psi=0$), and the maximum amplification
is $v/v_r = \cos^{-2} \psi_r = 1+(ST_s/2)^2$. Remembering that the amplification in
Fig.~\ref{fig:pre_fre_tb_g}(c) is proportional to $v^4$, we obtain
\beq
G\approx \left[  1+(ST_s/2)^2 \right]^4.
\la{eq:G}
\eeq
In essence, longer boxes have stronger two-dimensional bursts because the longer box
period gives the primary wavelength more time to amplify. Eq. \r{eq:G} is plotted as a
dashed line in Fig. \ref{fig:pre_fre_tb_g}(c), and captures reasonably well the rise of the
amplification. Note that a similar estimate holds for the higher two-dimensional Fourier
harmonics if $L_x$ is substituted by the wavelength $L_x/n_x$, and $T_s$ by $T_{sn}$. It is
harder to explain why the spikes cease to be amplified beyond $ST_s\approx 2$. The most
likely reason is nonlinearity. Fig.~\ref{fig:pre_fre_tb_g}(b) shows that the amplitude of
the spikes at $ST_s\approx 2$ is already about 10\% of the total, and it is probably
inconsistent to describe them with a linear theory. Also, the spectra in
Fig.~\ref{fig:traces}(d) show that the spectrum of the turbulent bursts begins to grow
beyond $ST\approx 3$--4, so that the spikes and the normal turbulence begin to interact
directly. In fact, the reason why the amplitude of the spikes does not decrease beyond $ST_s
\approx 2$ in Fig.~\ref{fig:pre_fre_tb_g}(b), even if the amplifications in
Fig.~\ref{fig:pre_fre_tb_g}(c) decrease, is that the Orr mechanism has stronger baseline
perturbations on which to work. It is conceivable that the background turbulence acts as a
eddy viscosity that prevents further growth of the Orr bursts.

\subsection{The numerical regeneration of linear Orr bursting}\label{sec:regen}

Although it is tempting to interpret the reconnection process in physical terms, it is clear
that it is an artifact of simulating the flow in a shear-periodic finite box. Homogeneous
shear flow has no implied temporal periodicity, even for flows that are periodic in $x$, and
the interactions described above do not exist. It is important to stress that only the
averaged correlations in Fig.~\ref{fig:movie2D} are smooth. The instantaneous flow has
smaller scales and is less ordered. It should also be noted that this two-dimensional
linear bursting is unrelated with the streak instability in the self-sustaining process of
shear turbulence,\cite{Waleffe1997} which is intrinsically three-dimensional.

To understand the numerical regeneration process we should go back to the sketch of
the implied Fourier grid in Fig.~\ref{fig:figgrid}(b). Note that the temporal evolution
of the Fourier modes in Eq.~\r{eq:sp_bctilde}, $k_y=\widetilde{k}_y-St k_x$, 
is the same as for those in the linearized RDT solution in Eq.~({\ref{eq:rdt_sol_1}}). 
Therefore, the amplitude of $v$ in each grid mode approximately satisfies the linearized RDT, 
which can be recast as $v \propto k_x^2/(k_x^2+k_y^2)$. 
The decrease in $k_y^2$ as the boundary condition approaches the top of the box cycle is 
the Orr amplification mechanism.

Consider the first streamwise Fourier mode, $n_x=1$. Near the bottom of the box
cycle, the two modes responsible for the largest-scale structures are those marked as
Q and Q' in Fig.~\ref{fig:figgrid}(b). The top mode, Q, whose wave-vector points
upwards $(k_y>0)$, is a backwards-tilted front such as the one in Fig.~\ref{fig:movie2D}(e).
The bottom mode Q' is a forward-tilted front. Statistically, both modes are similarly
forced by the nonlinearity, and their respective amplitudes are due to the linear processes
associated with their changing wavevector. For $t<T_s/2$, mode Q' is closest to
$k_y=0$ and is therefore the strongest of the two. As time increases, Q' moves away
from $k_y=0$ and weakens, while mode Q moves closer and strengthens. At $t=T_s/2$
both modes are equally intense on average. After that moment, the upper,
backwards-tilted mode predominates, and the structures flips backwards.

The linear Orr bursting is not restricted to long boxes. Whenever a particular
Fourier mode crosses the $k_y=0$ axis, the linear amplification process occurs to some
extent, even at moderate box aspect ratios. There are two time scales involved that
correspond to the two spectral components in Fig. \ref{fig:traces}(d). Two-dimensional
linear bursts have widths of the order $St\approx 2$, and are triggered by the boundary
conditions at longer intervals determined by the aspect ratio $\lambda_x/L_y$ of the
particular wavelength involved. The strongest resonance is for wavelengths of the order of
$L_x$, because they are the largest ones, and interact most strongly with their periodic
copies. They are triggered at intervals determined by the box aspect ration $L_x/L_y$, and
are the ones discussed in this section. Shorter harmonics also burst linearly, but they are
seldom two-dimensional, and their behavior tends to be dominated by nonlinearity. The direct
detection of the linear bursting in minimal channels is investigated in Ref.
\onlinecite{Jimenez2015linburst}. The result is that, once a burst is initiated by other causes,
its evolution can be predicted linearly over times of the order of 10--20\% of an eddy
turnover, corresponding to a tilting interval $\psi=-\pi/4$ to  $\psi=\pi/4$. Two-dimensional linear bursts become dominant in long boxes because there are
few nonlinear structures at the scale of the box length, and those that exist are only the
ones triggered by the two-dimensional boundary conditions.

\section{Comparison with other shear flows}\label{sec:ssp}

The previous section delimits the set of box aspect ratios in which the flow is as free as
possible from the artifacts of a finite simulation domain. The `acceptable' region is
summarized in Fig.~\ref{fig:symbols}(a) as the trapezoidal central region in which cases are
marked as circles (see also Table~\ref{tab:zones}). Note that very tall and long
boxes are defined by a further uncertain limit that has not been discussed up to now. Very
long and tall boxes satisfy all the above criteria, but the single example tested in that
region $(A_{xz}=A_{yz}=8$ at $Re_z=3000$ marked as a down-pointing triangle in
Fig.~\ref{fig:symbols}(a)) behaved strangely, with very deep three-dimensional bursts that
could not be described as Orr bursts. Those simulations are expensive to run. They are still
minimal in the spanwise direction, but otherwise very far from our desired range of minimal
flows, and this range was not pursued further.
  
A full discussion of what can be learned about the physics of shear turbulence from
such simulations is left to a later paper, but we briefly review in this section how
similar the statistically stationary HST is to other shear-dominated flows, particularly the
logarithmic layer of wall-bounded turbulence and the initial shearing of isotropic
turbulence.

\begin{table}[t]
\begin{center}
\caption{Parameters and statistics for the SS-HST simulations within the `acceptable'
range. Runs are labelled by their aspect ratios. Thus, L32 has $A_{xz}=3$ and
$A_{yz}=2$. $L_\varepsilon=(q^2/3)^{3/2}/\diss$ is the integral scale, and $\xi$ and $\zeta$
are the modified Lumley invariants, as explained in the text.
} 
\begin{tabular}{c*{11}{c}    }
\hline\hline 
 Run & $Re_z$ & $N_x, N_y, N_z$ &  $ST_\mathrm{stat}$ &  $Re_\lambda$ & $S^*$ & $\omega'/S$ & 
 $L_{\varepsilon}/L_z$ & $\xi$ &  $\zeta$\\
\hline 
  L32 &  2000 &  190,  192,   94 & 1706 &  47 & 6.82 & 5.36 & 0.41 & 0.092 & 0.104\\ 
  L34 &  2000 &  190,  384,   94 & 1581 &  48 & 6.97 & 5.30 & 0.42 & 0.095 & 0.105\\ 
  L38 &  2000 &  190,  768,   62 & 3177 &  48 & 7.02 & 5.34 & 0.43 & 0.097 & 0.106\\ 
  L44 &  2000 &  190,  384,   94 & 1570 &  48 & 7.10 & 5.28 & 0.43 & 0.093 & 0.100\\
  M32 & 12500 &  766,  512,  254 & 1029 & 105 & 7.53 & 10.9 & 0.38 & 0.101 & 0.108\\ 
  M34 & 12500 &  766, 1536,  382 & 1953 & 111 & 7.70 & 11.2 & 0.41 & 0.105 & 0.111\\ 
  H32 & 48000 & 2046, 2048, 1022 &  334 & 243 & 7.57 & 24.8 & 0.45 & 0.096 & 0.102\\
\hline\hline 
\end{tabular}
\label{table:HST_Axz3}
\end{center}
\end{table}

\begin{figure}[t]
\centering
\includegraphics[width=0.9\linewidth]{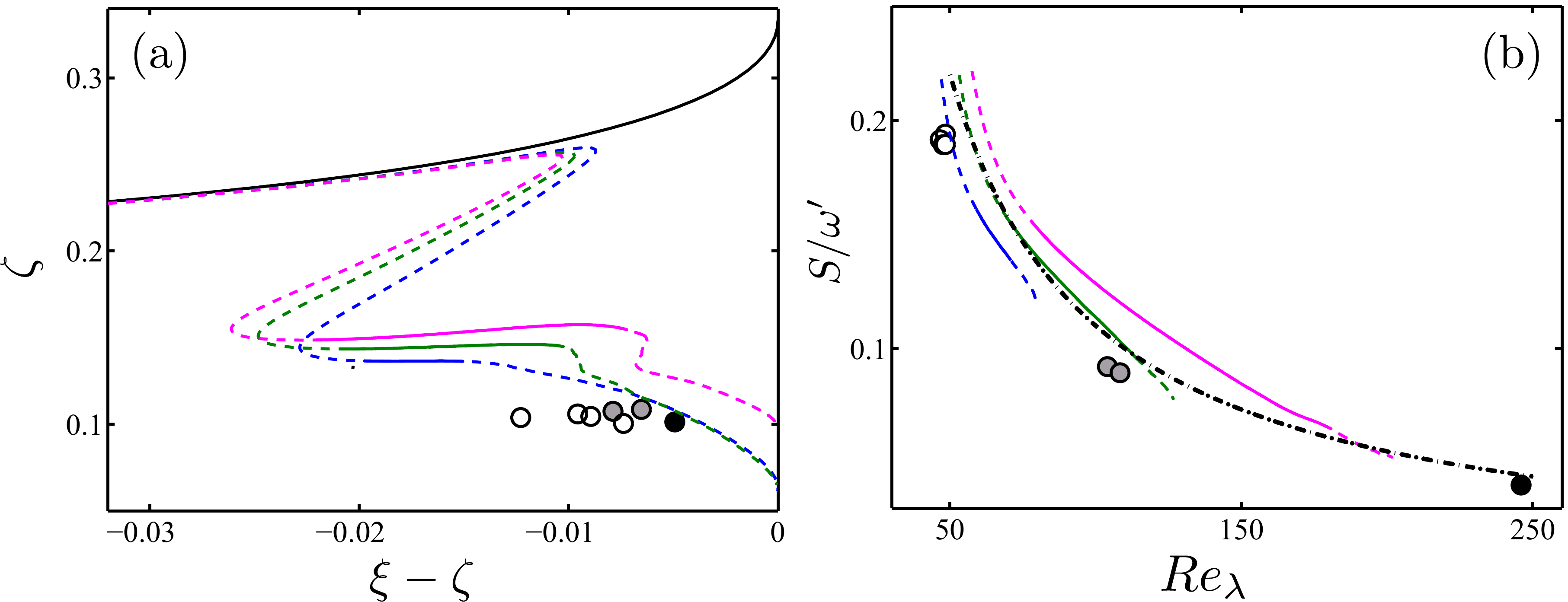}
\caption{%
(a) Modified Lumley invariants of the Reynolds-stress-anisotropy tensor.~\cite{pope00book}
The dark solid curve at the top of the figure, and the right-hand vertical axis are the realizability limits.
(b) $S/\omega'$ as a function of $Re_{\lambda}$. \chndot, $S Re_\lambda/\omega' =11$.
Lines are large-box simulations of channels at $Re_\tau=934$,
\cite{AlamoJimenez2004} 2003,\cite{HoyasJimenez2006} and 4200.
\cite{LozanoJimenez2014pof} Those in (a) go from the wall to the channel
center, and those in (b) go from $y^+=50$ to $y/h=0.4$, with $y$ increasing from left to
right. Higher Reynolds numbers correspond to longer lines. The solid segments represent the
logarithmic layer between $y^+=100$ and $y/h=0.2$. Circles are SS-HST in
Table~\ref{table:HST_Axz3}:
(white) L32, L34, L38, L44 from left to right;
(grey) M32, M34; (black) H32.
}
\label{fig:ll_so}
\end{figure}

Some of those comparisons have already been made in previous sections or
publications. The numerical parameters and some basic statistics of the acceptable
cases among our runs are summarized in Table~\ref{table:HST_Axz3}. When discussing
Fig.~\ref{fig:uvw_plust}, we mentioned that the velocity fluctuation intensities
are similar in SS-SHT and in the logarithmic layer. The correspondence improves if we
center our attention on the acceptable cases marked as circles, with the exception of
$v'^+$, which is stronger in the SS-HST than in channels ($v'^+\approx 1.2$--1.3 against
1.1). The relation between the different intensities is best expressed by the Lumley
invariants of the Reynolds-stress anisotropy tensor $b_{ij}=\bra u_i u_j \ket/ \bra
u_iu_i \ket-\delta_{ij}/3$, which are shown in Fig.~\ref{fig:ll_so}(a) in the
modified form, $6\zeta^2 =b_{ij}b_{ji}$, $6\xi^3=b_{ij}b_{jk}b_{ki}$ (see Ref.~\onlinecite{pope00book}). 
The figure includes results for large channels. They vary from
roughly axisymmetric turbulence, dominated by $u'$, near the wall, which is mapped in
the upper-left part of the figure, to approximate isotropy in the center of the
channel, mapped in the lower-right corner. The logarithmic layer is characterized by
an approximately constant second-order invariant, which is also where the SS-HST
tends to concentrate. The tendency of the latter to be more isotropic than the
channel reflects its stronger $v'$. The velocities tend to be more isotropic as the
Reynolds number increases, although there is a residual effect of the aspect ratios
for the SS-HST. Numerical values are given in Table~\ref{table:HST_Axz3}.
 
The strongest connection between the SS-HST and the logarithmic layer is that both are
approximately equilibrium flows in which the energy production equals dissipation.
The local energy balance can then be expressed as $u_\tau^2 S = \nu \omega'^2$ 
(see Ref.~\onlinecite{Jimenez2013nearwall}), and links quantities that are usually associated with
large and small scales. One example is the Corrsin parameter $S^*=Sq^2/\diss$,
which was shown in Ref.~\onlinecite{Jimenez2013nearwall} to be $O(10)$ in various shear flows,
including some of the present SS-HST simulations (see Table~\ref{table:HST_Axz3}).
Another example is the ratio $S/\omega'$, also listed in Table~\ref{table:HST_Axz3},
which is shown in Fig.~\ref{fig:ll_so}(b) as a function of $Re_\lambda$. The SS-HST and
the channels agree well. For both quantities, the energy equation can be manipulated
to give exact expressions: $S^*= {q^2}^+$ and $S Re_\lambda/\omega' = \sqrt{5/3}\,
{q^2}^+$. The agreement in Fig.~\ref{fig:ll_so}(b) and the approximate universality
of $S^*$ are then equivalent to the agreement of the intensities in Fig.~\ref{fig:uvw_plust}. 
Nonequilibrium flows, such as the initial shearing of isotropic turbulence,~\cite{RogersMoin1985,LeeKimMoin1990,Sarkar1995,IsazaCollins2009} 
do not agree either with the channels or with the SS-HST. Most of them do not even fall within
the ranges represented in Figs.~\ref{fig:ll_so}(a,b).
 
\begin{figure}[t]
\centering
\includegraphics[width=0.9\linewidth]{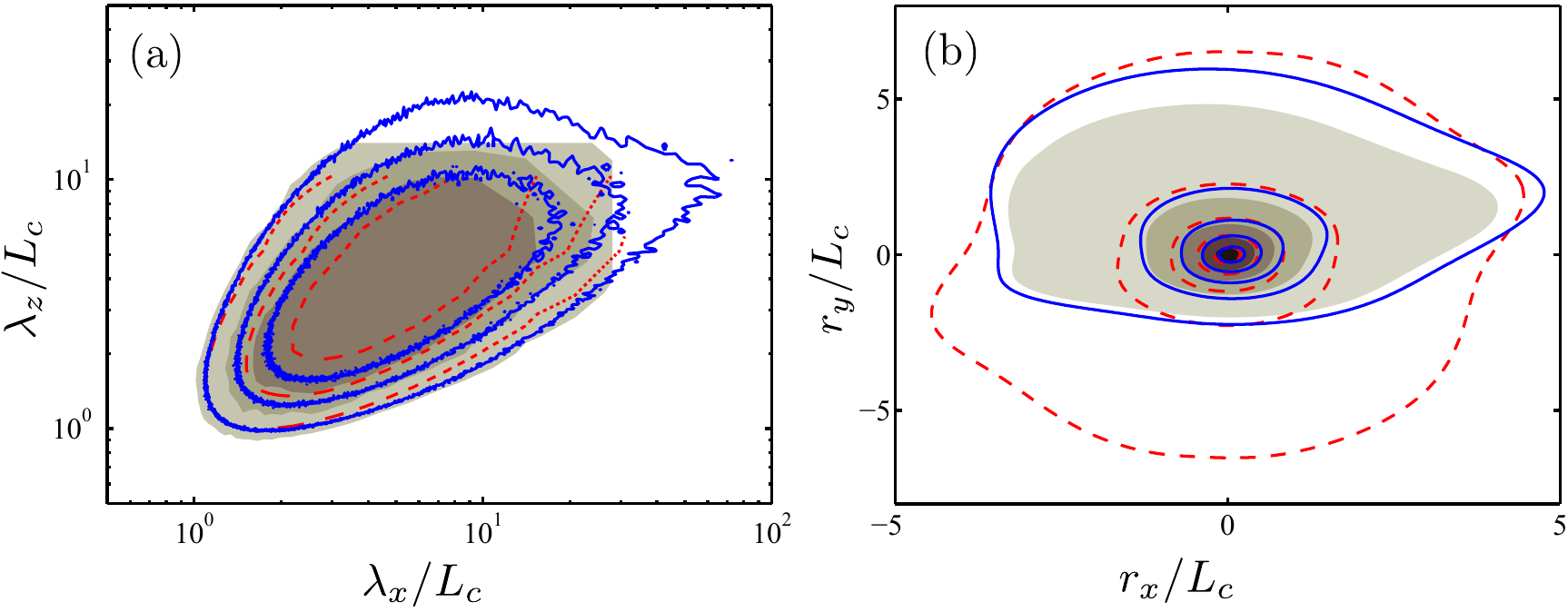}
\caption{
(a) Normalised premultiplied two-dimensional spectrum $k_xk_zE_{vv}/v'^2$. Contours
are [0.045:0.02:0.085].
(b) Two-point correlation function of the vertical velocity, $R_{vv}(r_x,r_y)$.  Contours
are [0.1:0.2:0.9].
Shaded contours are a minimal channel~\cite{FloresJimenez2010}: $Re_\tau=1840,\, L_x=\pi h/2,\, L_z=\pi h/4$; 
\solid, full channel~\cite{HoyasJimenez2006}: $Re_\tau=2003,\, L_x=8\pi h,\,
L_z=3\pi h$; \dashed, SS-HST (M32). The spectra for both
channels are at $y/h \approx 0.15$, where $Re_{\lambda}\approx 100$ as in the SS-HST
case. This is also the reference height for the correlations. Lengths are normalized
with the Corrsin scale $L_c=L_\varepsilon {(3/S^*)}^{3/2}$.
} 
\label{fig:prespec_vel}
\end{figure}

A more diagnostic property is the geometry of the structures.
Fig.~\ref{fig:prespec_vel}(a) shows the premultiplied two-dimensional spectrum of the
vertical velocity of the SS-HST (M32 in Table~\ref{table:HST_Axz3}), compared with those
of a large channel~\cite{HoyasJimenez2006} and of a minimal
one.~\cite{FloresJimenez2010} The wall distance of the channels $(y/h\approx 0.15)$
is chosen to match the three Reynolds numbers to $Re_\lambda\approx 100$,
and falls within the logarithmic layer. The boxes of the SS-HST simulation and of the
minimal channel are too small to capture the largest scales of $v$, but the agreement
of the three cases is surprisingly good. The small scales of the spectrum of $u$ and
$w$ (not shown) also agree well, although both quantities, which are typically larger
than $v$, are severely truncated in the two minimal boxes.

It is hard to define spectra in the non-periodic vertical direction, but
Fig.~\ref{fig:prespec_vel}(b) displays the two-point correlation function
$R_{vv}(r_x, r_y)$ for the same three cases, centered at the same location as in
Fig.~\ref{fig:prespec_vel}(a). The upper part of the correlations agrees relatively
well for the SS-HST and the large channel, even if the top of the outermost solid
contour is at $y/h\approx 0.45$, where the shear in the channel is three times weaker
than at the reference point. The upper part of the minimal channel agrees worse,
probably because its computational box is too narrow to allow good statistics above
$y\approx 0.3h$ (see Ref.~\onlinecite{FloresJimenez2010}).

The lower part of the correlations agree well for the two channels, but the SS-HST
extends much farther than any of the other two. The lower part of the outermost
contour of the two channels is very near the wall, but the SS-HST has no wall, and is
symmetric with respect to its center. It is interesting that, in spite of the large
differences in their lower part, the geometry of the upper part of the turbulent
structures is so similar between the three flows. The extra freedom of the vertical
velocity in the absence of the wall is probably responsible for the slightly higher
intensities of this velocity component in the SS-HST case. As in the spectra, the small
scales of $u$ and $w$ agree well, but the differences of the simulation boxes are too
large to allow a useful comparison between these larger structures.  

The length scale to be used in normalizing Fig.~\ref{fig:prespec_vel} is not immediately
obvious. Viscous wall units are not relevant for the SS-HST, and it is indeed found that
they collapse the correlations poorly. The integral length $L_\varepsilon$ works better but,
after some experimentation, the best normalization is found to be the Corrsin scale. This is
probably physically relevant. That length was introduced in Ref.~\onlinecite{Corrsin1958} as
the limit for small-scale isotropy in shear flows, and it is roughly the scale of the
smallest tangential Reynolds stresses. It can be interpreted as the energy-injection scale
for shear flows, and it is related to the integral scale by $L_c \equiv
(\diss/S^3)^{1/2} =L_\varepsilon (3/S^*)^{3/2}$. We have mentioned that $S^*$ is
fairly similar in SS-HST and channels, but the differences are enough to substantially
improve the collapse of the correlations and of the spectra.

\begin{figure}[t]
\centering
\includegraphics[width=0.8\linewidth]{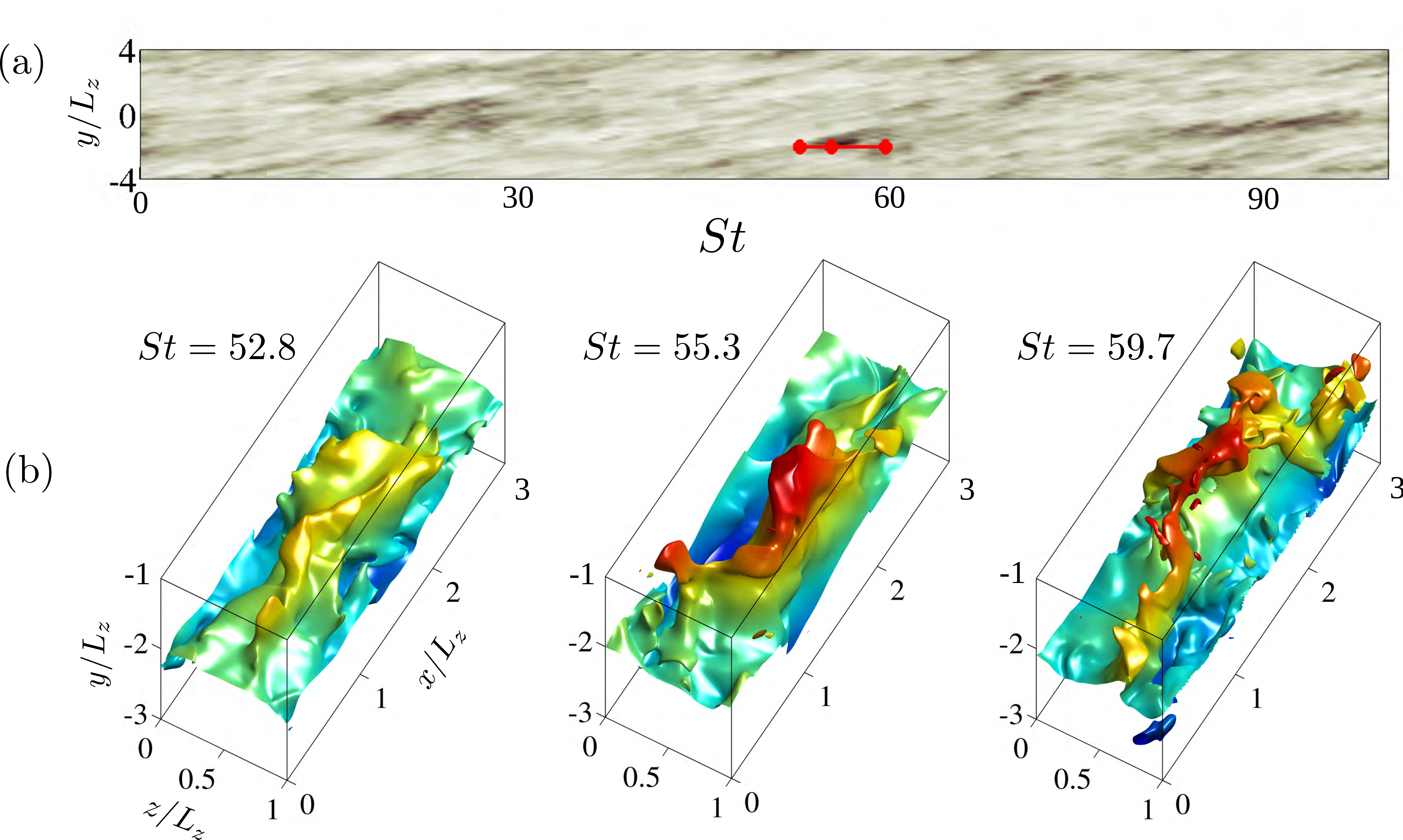}
\caption{%
(a) Temporal evolution of $-\bra uv \ket_{xz}$, conditionally-averaged for $v>0$ over wall-parallel planes. 
Strong bursting events are represented by darker gray. 
(b) Isosurface of the total streamwise velocity $u+Sy=-2SL_z$ at three moments of a bursting event marked 
by red circles in (a). Only one fourth of the vertical domain is shown, centered at
$y/L_z=-2$. The isosurfaces are coloured by the mean streamwise velocity $Sy$.
Case L38.
} 
\label{fig:q2q4}
\end{figure}

We finally compare the bursting behavior of SS-HST and wall-turbulence. Fig.~\ref{fig:q2q4} 
shows the evolution of the streamwise-velocity streak in SS-HST during a
burst. Comparison with the logarithmic layer of a minimal channel in Fig.~3 of
Ref.~\onlinecite{FloresJimenez2010} strongly suggests that the two phenomena are related.

\begin{figure}[t]
\centering
\includegraphics[width=0.85\linewidth]{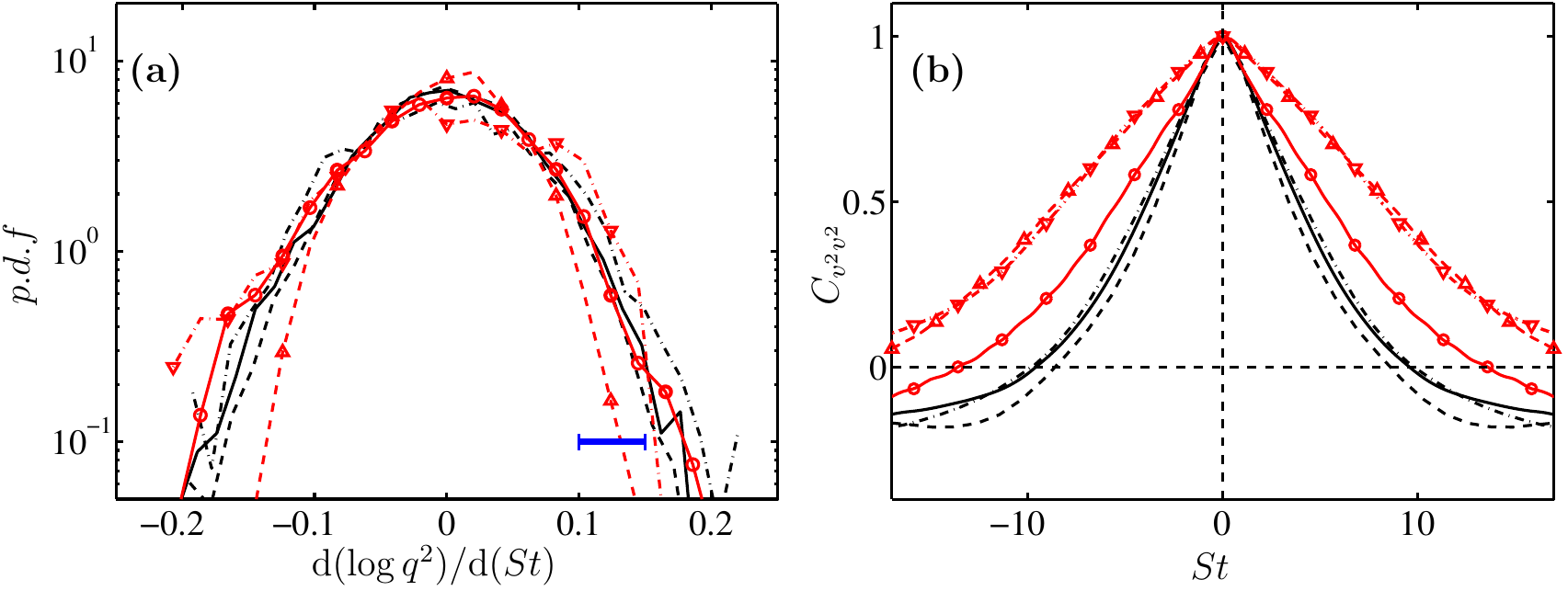}
\caption{%
(a) P.d.f. of the instantaneous growth rates of the kinetic energy. Lines without
symbols are minimal channels~\cite{FloresJimenez2010} in the band $y/L_z\approx 0.13$--0.25:
\solid, $(Re_\tau, Re_\lambda)=(1830, 125)$;  
\chndot, $(1700, 80)$;   
\dashed, $(950, 90)$.
Lines with symbols are SS-HST from Table~\ref{table:HST_Axz3}:
$\circ$, L32; $\vartriangle$, M32; $\triangledown$, H32.
The horizontal bar is the range of growth rates for weak initial shearing of isotropic turbulence.
(b) Temporal autocorrelation function of $\bra v^2\ket_V$. The channels are as in (a), but the
SS-HST are now: $\circ$, L32; $\vartriangle$, M32; $\triangledown$, M34.
} 
\label{fig:bursting}
\end{figure}

Previous investigations of SS-HST have suggested that the growth phase of bursts is
qualitatively similar to the shearing of initially isotropic
turbulence.~\cite{Pumir1996,Gualtieri2002} Fig.~\ref{fig:bursting}(a) provides some
quantitative evidence and qualifications. It contains probability density functions~(p.d.f.)
for the logarithmic growth rate of the kinetic energy, ${\it \Lambda}=\dr(\log q^2)/\dr
(St)$, in SS-HST and minimal channels. This quantity has been widely discussed for the
shearing of initially isotropic turbulence, and is believed to settle
asymptotically to ${\it \Lambda}\approx 0.1$--0.15 in weakly sheared cases $(S^\ast \lesssim
10)$,~\cite{RogersMoin1985,Sarkar1995} and to somewhat higher values, ${\it \Lambda}\approx
0.2$, in strongly sheared ones.~\cite{LeeKimMoin1990,IsazaCollins2009} In statistically
stationary flows, the mean value of ${\it \Lambda}$ vanishes, but the distribution of its
instantaneous values can be used as a measure of how fast bursts grow during their
generation phase, or otherwise decay.

The p.d.f.s for the SS-HST flows in Fig.~\ref{fig:bursting}(a) are for three cases
with the same aspect ratios, $A_{xz}=3, A_{yz}=2$, but different Reynolds numbers.
Those for the minimal channels are compiled over the range of height $y/L_z
\approx 0.13$--0.25, which is known to be roughly minimal for the largest structures of the
logarithmic layer.~\cite{FloresJimenez2010} By changing $Re_\tau$ and the width $L_z$
of the simulation box, the Taylor-microscale Reynolds number of the three channels are also
made different. The agreement of the different p.d.f.s again supports that the
dynamics of bursting in minimal channels and in SS-HST is similar. The range of initial 
growth rates found in weakly sheared turbulence is marked in Fig.~\ref{fig:bursting}(a) 
by a horizontal bar. It spans the near-tail of the
p.d.f.s, consistent with the interpretation mentioned above that the growth of
individual bursts is similar to the weak shearing of initially isotropic turbulence.
The values of $S^*$ in Table~\ref{table:HST_Axz3} confirm that SS-HST is in the
weakly sheared regime.

The p.d.f. of ${\it \Lambda}$ depends on the size of the computational box. 
Taller SS-HST boxes or channel heights closer to the wall produce narrower distributions. 
The common property of the boxes in Fig.~\ref{fig:bursting}(a) is that they are minimal, 
in the sense that they are expected to contain a single large structure on average. 
Taller boxes can be expected to contain several structures, evolving roughly independently. 
The same is true of near-wall planes in the channel, where several
small near-wall structures fit within the simulation box. In those cases, the energy of
the individual structure add to the total energy and the standard deviation of the
total growth rate should decrease roughly as the square root of the expected number
of structures. This is confirmed by the SS-HST simulations. The standard deviation of
${\it \Lambda}$ in cases L32, L34 and L38, expected to contain one, two and four structures,
are in the ratio 1.96, 1.39 and 1, which are almost exactly powers of $\sqrt{2}$.

The situation is the opposite for the temporal correlation functions used to compute the
bursting period in Fig.~\ref{fig:ene_10}(b), a few of which are displayed in
Fig.~\ref{fig:bursting}(b). It was shown in Ref.~\onlinecite{Jimenez2013nearwall} that the
width of the temporal correlation of $v$ is of the same order of magnitude in both flows,
but Fig.~\ref{fig:ene_10}(b) shows that the correlations of the SS-HST are somewhat wider
than in the channels. They also widen with increasing Reynolds number. Two of the SS-HST
simulations in Fig.~\ref{fig:ene_10}(b) (L32 and M32) have the same aspect ratios and
different $Re_\lambda$. The higher Reynolds number is substantially wider than the lower
one. Two other simulations (M32 and M34) have similar Reynolds numbers and different aspect
ratios, and their correlations agree. It can be shown that the wider correlations are due to
the different behavior of the higher Fourier modes. If these are filtered out, 
the temporal correlations of the longest Fourier mode $(n_x=1)$ collapse for all Reynolds numbers.

The width of individual bursts, $St\approx 20$, is thus determined by the linear
transient amplification of random nonlinear perturbations, as already shown in Refs.
\onlinecite{Jimenez2013howlinear,Jimenez2013nearwall}. On the other hand, the time between
consecutive bursts is determined by how often these particular initial conditions are
generated, and is beyond the scope of the present paper. The long-term behavior of sheared
turbulence is a controversial subject,~\cite{CambonScott1999} but the wider consensus seems
to be that it grows indefinitely, both in intensity and in length
scale.~\cite{RogersMoin1985,TavoularisKarnik1989,IsazaCollins2009} In simulations, the
growth of the length scale will eventually interfere with any finite computational box, and
simulations of the initial shearing have traditionally been discontinued at that moment. It
is generally believed that the periodic bursting of statistically stationary HST is due to
the periodic filling of the computational box by the growing length scales. We have seen in
Sec.~\ref{sec:regen} a particular numerical artifact leading to the regeneration of
linear bursts, and there is circumstantial evidence from our simulations that some of the
length scales reach the box limit before the collapse of the intensities. On the other hand,
Fig.~\ref{fig:ene_10}(b) shows that the temporal width of the bursts is relatively
independent of the box dimensions within certain limits, and that it scales well with the
shear. A similar argument about length scales could be made about channels, although
inhomogeneity and the wall may play a role similar to geometry in that case. However, we
have just shown that channels share many properties with SS-HST, and it was shown in
Ref.~\onlinecite{Jimenez2013nearwall} that the bursting period of minimal channels, which
are geometrically constrained, is similar to the lifetime of individual structures in larger
simulations,~\cite{LozanoJimenez2014JFM} which are not. Both are of the same order as the
bursting width of SS-HST (Fig.~\ref{fig:bursting}b). In summary, it appears that bursting is
a property of shear flows in general, not linked to the presence of a wall, but its
properties require further work.

\section{Conclusions}\label{sec:conc}

We have performed direct numerical simulations of homogeneous shear turbulence (HST)
to explore the parameter range of statistically stationary HST (SS-HST). Our code uses a
shear-periodic boundary condition in the vertical direction that requires no periodic
remeshings, and that is implemented directly on the compact finite difference. 
The other directions are spectral. Validations collected in Appendix C of the
supplementary material~\cite{Supplemental1} confirm that the code maintains its designed accuracy, and
reproduces well previous simulations. In all those cases, our small-scale statistics
are slightly higher than those in remeshing codes using the same Fourier modes,\cite{Rogallo1981}
probably because the lack of remeshing prevents the loss of some enstrophy. We have
given a simple theoretical analysis for why that should be so.

The statistics of SS-HST are strongly dependent on the geometry of the computational
box, represented by its two aspect ratios, $A_{xz}=L_x/L_z$ and $A_{yz}=L_y/L_z$. We
have shown that the relevant length scale is the spanwise width of the box, $L_z$,
and that the velocity scale is $SL_z$. The relevant Reynolds number is therefore
$Re_z\equiv SL_z^2/\nu$, but the characteristics of the largest-scale motions in
SS-HST are found to be fairly independent of $Re_z$. It is interesting that $L_z$ is
also the dimension that determines the range of valid wall distances in minimal
turbulent channels.~\cite{FloresJimenez2010} Since it is believed that the large
scales of theoretical homogeneous shear flow have no characteristic length, flows
simulated over long enough times tend to `fill' any computational box, and long-term
simulations of SS-HST are always `minimal' in the sense that there are only a few
largest-scale structures in the computational box.

The empirical evidence is that the effect of the geometry can be reduced by ensuring
that box dimensions other than $L_z$ do not constrain further these minimal
structures. The limits identified are $A_{xz}>2$ and $A_{yz}>1$. A similar argument
can be made for minimal channels,~\cite{FloresJimenez2010} even though `natural'
structures are always influenced by the wall in that case. The effect of minimal
boxes is to override that constraint when the box becomes too narrow far enough from
the wall. It follows that the limit in which the channel is `just minimal' $(y\approx
L_z/3)$ (see Ref.~\onlinecite{FloresJimenez2010}) should have similar properties to
`just minimal' SS-HST.

That turns out to be approximately true. The one-point statistics of SS-HST in the
suitable range of the aspect ratios agree surprisingly well with those of the
logarithmic layer in turbulent channel flows, particularly when scaled with the
friction velocity derived from the measured Reynolds stresses. The same is true for
the wall-parallel spectra of the wall-normal velocity, although the length scales of
the other two velocity components are typically too large to be fruitfully compared
either with SS-HST or with minimal channels. The wall-normal spatial correlation of
$v$ also agrees well with channels and boundary layers, but only in the direction
away from the wall. The correlations of channels in the direction of the wall are
limited by impermeability, but those of the SS-HST are not, and extend symmetrically
downwards. It is interesting that even this strong difference does not influence the
top part of the correlation.

An interesting limit of SS-HST that does not appear to exist in minimal channels is that of
very long boxes $(A_{xz} \ge 2A_{yz})$. Their bursting is dominated by a two-dimensional
numerical regeneration process associated with the interaction between shear-periodic copies
of the numerical box. It can be treated analytically to a large extent, but appears to be
unrelated to physics.

A common characteristic of SS-HST and wall-bounded turbulence is quasi-periodic bursting,
and we have shown that it shares many common characteristics in both flows. Besides strong
similarities of the flow fields, the lifetime of individual bursts, defined from
the temporal autocorrelation function, scales with the shear in both cases as $ST_b \approx
20$. In contrast with the numerical two-dimensional bursts described in the previous
paragraph, these ones involve the quasi-simultaneous growth of the three velocity
components, and presumably originate from the linear amplification of three-dimensional
`dangerous' initial conditions~\cite{schm07,Jimenez2013howlinear} randomly found as parts of
the nonlinear turbulent field. The probability distribution of the growth rates of the
intensities suggests that the amplification phase of the bursts is similar to the weak
shearing of initially isotropic turbulence, where the generation of
initial conditions is presumably the same.  In general, it is concluded that the similarities
between SS-HST and other shear flows, particularly with the logarithmic layer of wall
turbulence, make it a promising system in which to study shear turbulence in general.

\section*{Acknowledgements}
This work was partially funded by the ERC Multiflow project. S. Dong was supported by
the China Scholarship Council. The authors gratefully acknowledge the computing time
granted by the Prace European initiative on the supercomputer JUQUEEN at the
J\"ulich Supercomputing Center, and by the Red Espa\~nola de Supercomputaci\'on on
Marenostrum at the Barcelona Supercomputing Centre. We are grateful to R.R. Kerswell 
for a careful critique of the original manuscript.
 
\clearpage
  \newcommand{\fomy}{\widehat{\omega_y} }
  \newcommand{\fphi}{\widehat{\phi} }
\appendix
\section{Three-step fully-explicit Runge--Kutta with analytical integration of the
shear convective terms}\label{appendixA}

Applying the Fourier transform to the governing equations (Eqs.~(1,2) in the manuscript), 
we have  in general, 
\begin{eqnarray}
   \pdif{f}{t} + \mathrm{i} k_x Sy f =  R(t,f),	\label{eq:fff}
\end{eqnarray}
where $f$ represent any of~$\fomy$, $\fphi$, $\langle u \rangle_{xz}$, or $\langle w
\rangle_{xz}$. We analytically absorb the linear shear convective term $\mathrm{i}
k_x Sy f$ in Eq.~(\ref{eq:fff}) by multiplying it by the integrating factor
$\exp(\mathrm{i} k_x S y t)$,
\begin{eqnarray}
  \pdif{(e^{\mathrm{i} k_x S y t} f )}{t}  &=& e^{\mathrm{i} k_x S y t} R(t,f).
\end{eqnarray}
The semi-discrete form of the three-step fully-explicit Runge--Kutta scheme~\cite{Spalart1991} 
to advance from $f(t)$ to $f(t+\Delta t)$ leads to,
\begin{eqnarray}
f^\ast  &=&  f +\gamma_1 \Delta t R(t,f), \notag \\
f_1  &=&  e^{- \mathrm{i} k_x S y c_1 \Delta t} f^\ast, \label{eq:mapping1}\\
R_1 &=& e^{- \mathrm{i} k_x S y c_1 \Delta t}  R(t,f), \\
f^{\ast\ast} &=& f_1 + \gamma_2 \Delta t R(t+c_1 \Delta t, f_1) + \zeta_1 \Delta t  R_1, \notag \\
f_2 &=&  e^{-\mathrm{i} k_x S y (c_2-c_1) \Delta t}f^{\ast\ast},  \\
R_2 &=& e^{- \mathrm{i} k_x S y (c_2-c_1) \Delta t}  R(t+c_1 \Delta t, f_1), \\
f^{\ast\ast\ast}  &=& f_2 +\gamma_3 \Delta t R(t+c_2 \Delta t, f_2)+ \zeta_2 \Delta t R_2, \notag\\
f_3 &=&  e^{- \mathrm{i} k_x S y (c_3 - c_2) \Delta t} f^{\ast\ast\ast}, \label{eq:mapping5}
\end{eqnarray}
where $f^\ast$, $f^{\ast\ast}$, $f^{\ast\ast\ast}$, $f_1$ and $f_2$ represent the
intermediate variables at each Runge-Kutta sub-step, $i$=\{1,2,3\}, and
$f_3=f(t+\Delta t)$ corresponds to the next time step. The coefficients are:
\begin{eqnarray}
\gamma_i =\left\{ \frac{8}{15},~ \frac{5}{12},~ \frac{3}{4}  \right\},~~
\zeta_i =\left\{-\frac{17}{60},~-\frac{5}{12}  \right\},~~
c_i =\left\{ \frac{8}{15},~ \frac{2}{3}, ~1  \right\}.
\end{eqnarray}
This scheme is third-order consistent. 
The additional operations over a traditional integrator are the five `unmapping'
multiplications in Eqs.~(\ref{eq:mapping1})-(\ref{eq:mapping5}) by $\exp[-\mathrm{i}
k_x S y (c_{i+1} - c_{i})\Delta t]$ ($c_0=0$). In our simulations, the cost of
mapping is roughly 10\% of the total, but it reduces the advective CFL by the ratio
$2 u'/SL_y$, which can be considerable, especially for tall computational boxes, $A_{yz}>1$.

A semi-implicit scheme for the viscous term could also be used (e.g.,
Ref.~\onlinecite{Spalart1991}), but it is useful only at very low Reynolds numbers (roughly
$Re_z < 1000$ in the present case) for which the viscous CFL leads to a smaller time
step than the advective one.

\section{Compact finite differences with a shear-periodic boundary condition}\label{appendixB}

In order to compute derivatives in the vertical direction ($y$), we use a
compact-finite-differences scheme~\cite{Lele1992} based on a seven-point stencil with
6th- and 8th-order resolution accuracy for the first and second derivative, respectively.
Exact spectral behavior is enforced at the wavenumbers $k\Delta y/\pi=0.5, 0.7, 0.9$
for the first derivative, and $k\Delta y/\pi=0.5, 0.9$ for the second one. The modified
wavenumber $k^\prime\Delta y$ estimated by Fourier analysis for the compact finite
differences described in this section stays close to the exact differentiation over a
range of wavenumbers $k^\prime\Delta y \le 2.5$, which is used for the estimation of the
resolution requirements of the DNS. The consistency errors, $\varepsilon_1 \equiv
|k^\prime - k|/k$ for the first derivative, and $\varepsilon_2 \equiv |k^{\prime 2} -
k^2|/k^2$ for the second one, are $\varepsilon_1 \approx 0.006$ and $\varepsilon_2
\approx 0.005$, respectively, at the adopted resolving efficiency $k\Delta y=2.5$.

The discretized form of the $n$-th derivative of $f(y_j) \approx F_j$ in the $y$-direction,
where $ y_j \equiv (j - 1) L_y/N - L_y/2,~j=1, ..., N $, is written as
\begin{eqnarray}\label{eq:compact}
  B F^{(n)} = A F,
\end{eqnarray}
where $ F^{(n)}$ represents the $n$-th derivative of $F$.
Assuming an even derivative, the structure of the matrix $B$ is
\begin{eqnarray}\label{Bmat}
B =  \left(%
\begin{array}{ccccc ccccc}
\delta &\alpha &\beta &\gamma &0 & \cdots &0& \gamma^{\prime\ast} & \beta^{\prime\ast} & \alpha^{\prime\ast} \\
\alpha & \delta &\alpha &\beta &\gamma &0 & \cdots &0& \gamma^{\prime\ast} & \beta^{\prime\ast} \\
\beta  & \alpha &\delta &\alpha &\beta &\gamma &0 & \cdots &0& \gamma^{\prime\ast} \\
\gamma & \beta  & \alpha &\delta &\alpha &\beta &\gamma &0 & \cdots &0 \\
0&\gamma & \beta  & \alpha &\delta &\alpha &\beta &\gamma &0 & \cdots \\
\vdots &  & \ddots  &  & &\ddots & & & \ddots& \vdots \\
0 & \cdots &0 & \gamma & \beta  & \alpha &\delta &\alpha &\beta &\gamma \\
\gamma^{\prime} & 0 & \cdots &0 & \gamma & \beta &\alpha &\delta &\alpha &\beta \\
\beta^{\prime} &\gamma^{\prime} & 0 & \cdots &0 & \gamma & \beta  & \alpha &\delta &\alpha \\
\alpha^{\prime} &\beta^{\prime} &\gamma^{\prime} & 0 & \cdots &0 & \gamma & \beta  & \alpha &\delta
\end{array}
 \right),
\end{eqnarray}
and $A$ has the same structure of non-zero entries, with different coefficients.
 Note that
$\alpha$, $\beta$, $\gamma$ and $\delta$ are constant real values. The application of the
shear-periodic boundary condition 
\beq
F_j(t, k_x, k_z) = F_{j+N}(t, k_x, k_z) \exp[\ii k_x S L_y t],
\label{eq:sp_bc_app}
\eeq
to the compact finite difference matrices appears in its off-band-diagonal elements, which
are complex $\alpha^\prime$, $\beta^\prime$, $\gamma^\prime$, and their complex conjugates
$\alpha^{\prime\ast}$, $\beta^{\prime\ast}$, and $\gamma^{\prime\ast}$. Specifically,
$\alpha^\prime = \alpha \exp[-\ii k_x S L_y t]$, etc., which is used to substitute off-grid
elements by their shifted copies near the opposite boundary. I.e., $F_0 = F_N \exp[\ii k_x
\Delta U t]$, $F_{N+1} = F_1 \exp[-\ii k_x \Delta U t]$, etc., where $\Delta U = SL_y$ is
the mean velocity difference between the two boundaries. Therefore, $A$ and $B$ are
time-dependent Hermitian and need to be updated at each Runge--Kutta sub-step.
Odd derivatives are handled similarly with a skew-Hermitian $A$.

The linear system~(\ref{eq:compact}) is directly solved by applying the modified
Cholesky decomposition $B=LDL^{\ast}$,
\begin{eqnarray}\label{Lmat}
L = \left(%
\begin{array}{ccccc ccccc}
1   & 0 & &   & & \cdots && & &0 \\
a_{2} & 1 & 0 &   & & & \cdots && & \\
b_3 & a_3 & 1 & 0 & & & & & & \\
c_4 & b_4 & a_4 & 1 & 0 & & & & & \\
0   & c_5 & b_5 & a_5 &1 & 0 & & & & \\
\vdots &  & \ddots  &  & &\ddots & & & \ddots& \vdots \\
0 & \cdots &0 & c_{N-3} & b_{N-3} & a_{N-3} &1 & 0 & & \\
e_1 & e_2 & \cdots & & e_{N-5} & e_{N-4} & e_{N-3} &1 & 0 & \\
f_1 & f_2 & \cdots & & \cdots & f_{N-4} & f_{N-3} & f_{N-2} &1 & 0 \\
g_1 & g_2 & \cdots & & & \cdots & g_{N-3} & g_{N-2} & g_{N-1} &1 \\
\end{array}
 \right), 
\end{eqnarray}
and 
\begin{eqnarray}\label{Dmat}
D = \left( \begin{array}{ccccc} 
d_1&&&&0 \\
&\ddots&&& \\
&&d_{i}&& \\
&&&\ddots& \\
0&&& &d_N
\end{array} \right).
\end{eqnarray}
The modification in the time-marching in DNS is done only for the three complex lines
$e_i$, $f_i$ and $g_i$ for the matrix $L$, and their complex-conjugates for $L^\ast$.
Note that the band-diagonal elements $a_i$, $b_i$, $c_i$ and the diagonal elements
$d_i$ are real constant.

The one-dimensional Helmholtz equation, expressed generally as $F^{(2)} +
\lambda F = R_f$ (where $\lambda$ is real) leads to a linear system $(A + \lambda B) F = B
R_f$, which can be solved for $F$ by applying the modified Cholesky
decomposition of the Hermitian operator $(A + \lambda B)$.

\section{Validations}
\subsection{Rapid distortion theory}\label{sec:valid}

When the velocity gradient fluctuations are small with respect to the mean shear, the
Navier--Stokes equations can be linearized. In this `rapid distortion' limit,
\beq\label{eq:RDT3}
\partial_t \uvec = - (\Uvec \cdot \nabla) \uvec - (\uvec \cdot \nabla) \Uvec - \nabla
p + \nu \nabla^2 \uvec,
\eeq
where $\uvec$ and $p$ are infinitesimal. Note that when these linearized equations are written in terms of the variables $\nabla^2 v$ and $\omega_y$, they reduce
to the classical Orr--Sommerfeld and Squire equations, respectively. 

For individual Fourier modes in a  pure shear, the velocities can be expressed as
$\uvec = \sum_{m} \widehat{\uvec}(t) \exp[\ii k_m (t) x_m],\, m=x,y,z$, 
and Eq.~(\ref{eq:RDT3}) becomes
\begin{eqnarray}\label{eq:RDT4}
  \p_t \hu &=& (k_{0x}^2 - k_{0z}^2 - k_y^2) S \hv/|\kvec|^2 - \nu |\kvec|^2 \hu, \nonumber \\
  \p_t \hv &=& 2 k_{0x} k_y S \hv/|\kvec|^2 - \nu |\kvec|^2 \hv, \\
  \p_t \hw &=& 2 k_{0x} k_{0z} S \hv/|\kvec|^2 - \nu |\kvec|^2 \hw, \nonumber \\
  k_y & =&  k_{0y}-S k_{0x} t,  \nonumber
\end{eqnarray}
where $\kvec_0 = (k_{0x},k_{0y},k_{0z})$ and $\kvec = (k_{0x}, k_y, k_{0z})$ are the
initial and time-evolving wave vectors. These equations can be solved analytically
\cite{Moffatt1967,Townsend1976book,far_ioa93a}, and are used to exercise
the linear parts of the code.

\begin{figure}[t]
\centering
\begin{minipage}{.80\linewidth} 
   \includegraphics[width=1.0\linewidth,clip]{\figpath 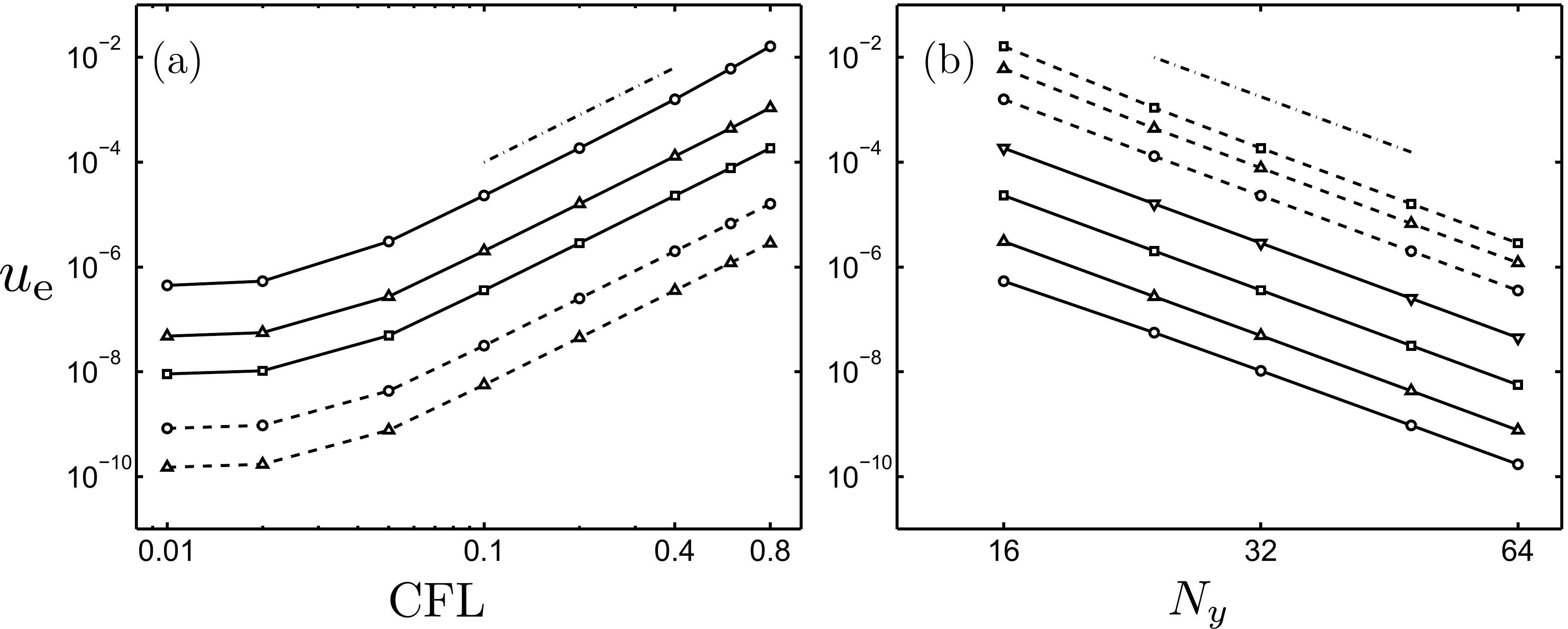}
\end{minipage}
\caption{Relative error $u_{e}$ for the streamwise velocity, compared with the
corresponding linear RDT solution.
(a) For different grids in $y$, as a function of the CFL. \solidcirc, $N_y=16$;
\solidtri 24; \solidsquare, 32; \dashedcirc, 48; \dashedtri, 64. In all cases, $N_x
= N_z = 18$. The chaindotted line has slope 3. For other parameters, see text.
(b) As in (a), as a function of $N_y$. \solidcirc, CFL$=0.02$; \solidtri, 0.05; \solidsquare, 
 0.1; \solidtridown, 0.2; \dashedcirc, 0.4; \dashedtri, 0.6; \dashedsquare, 0.8. 
The chaindotted line has slope $-6$.
}
\label{fig:rdt:CFL_u} 
\end{figure}
\begin{figure}[t]
\centering
\includegraphics[width=.72\linewidth,clip]{\figpath 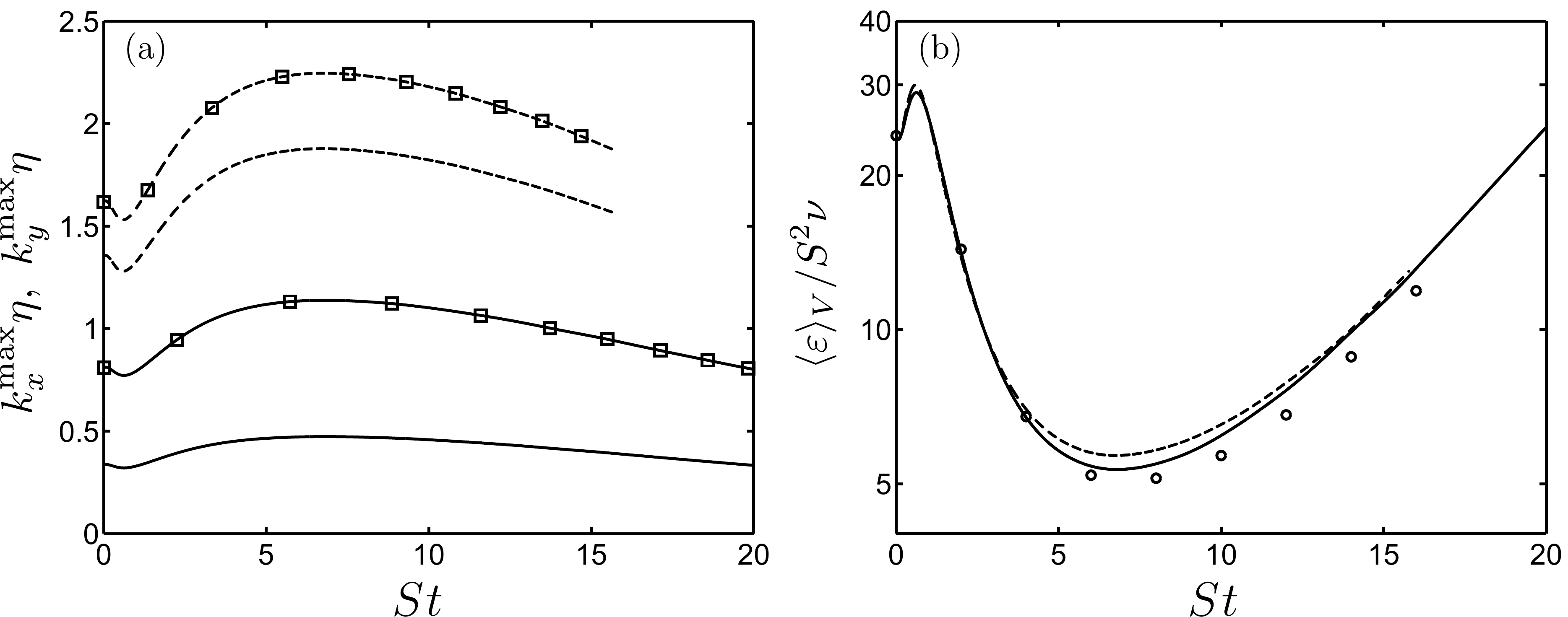}%
  \caption{%
Effect of the grid resolution. Case HOM23U.~\cite{RogersMoin1985} (a) The time
evolution of the effective resolutions, $\itbold{k}_\mathrm{max} \eta(t)$. Lines with
symbols are $k_x \eta$; without symbols are $k_y \eta$.
(b) Evolution of the energy dissipation rate. $\circ$, HOM23U.~\cite{RogersMoin1985}
In both figures, \dashed, fine grid, ($510, 384, 254$); \solid, coarse grid
($126, 192, 126$).
}
  \label{fig:validation1_fine} 
\end{figure}  
\begin{figure}[t]
  \centering
    \includegraphics[width=0.95\linewidth,clip]
    {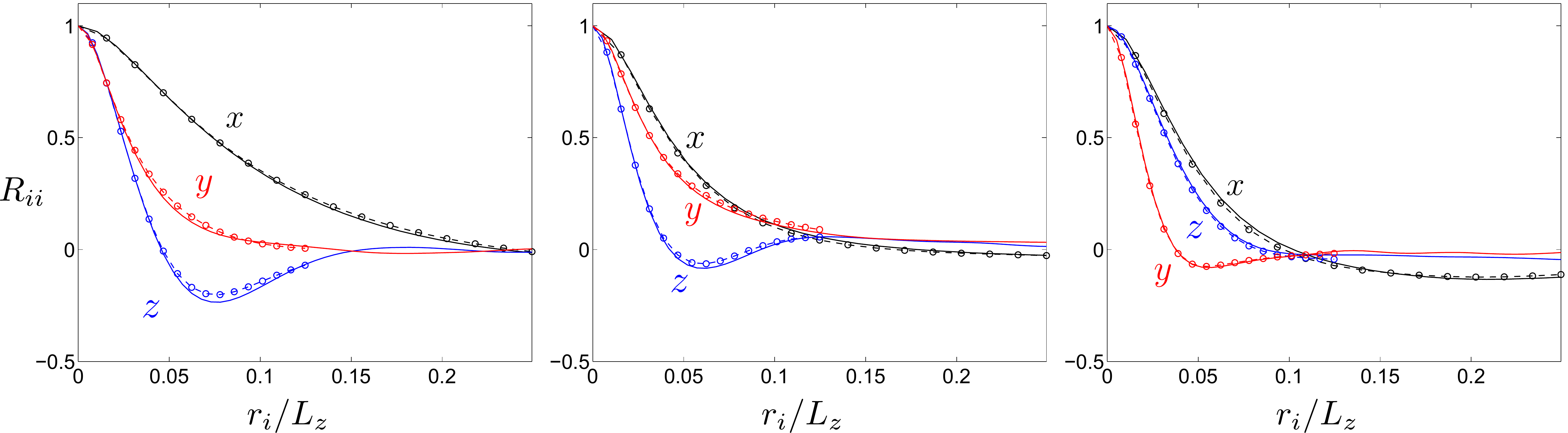}
  \caption{ Auto-correlation functions for streamwise (a), vertical (b),
   and spanwise (c) velocities 
   along $x$- (black), $y$- (red), $z$- (blue).
   \solid, present DNS ($192\times192\times192$ with CFL=0.6);
   \dashedcirc, HOM23U in Ref.~\onlinecite{AGARD}, from Rogers {\it et al.}~\cite{RogersMoin1985}.
   $ St = 10.0 $.
   }
  \label{fig:validation_corr} 
\end{figure} 

Fig.~\ref{fig:rdt:CFL_u} shows the time-integrated relative error, $u_{e}^2 =
t_l^{-1}\int_{0}^{t_l} \bra |u - u_\mathrm{RDT}|^2 \ket_V/ |\uvec_0|^2 \dd t$,
between the streamwise velocity in the present DNS and the corresponding RDT
solution. Both in DNS and RDT, $S = 1$ and $Re_z= 4\times 10^4$. The initial
condition is a sine wave, $\uvec_0 = (0, v_0, 0)\sin(k_{0x} x +k_{0z} z)$, with
initial wavenumbers $\kvec_0 L_z = (0.5, 0, 1)$. In the DNS, the initial
amplitude is $|{v}_0|/(S L_z) = 10^{-3}$, and the box aspect ratios are
$(A_{xz},A_{yz}) = (2, 1/\pi)$, so that the box always contains a single wavelength
in the horizontal plane. The simulations run from $t=0$ to $St_l=15$, by which time
the magnitude of the vertical wavenumber reaches $|k_y \Delta y| \approx 1$ for
$N_y=16$. This is a typical minimum resolution in later turbulence simulations. 
Note that 
the cases in Fig.~\ref{fig:rdt:CFL_u} imply run times of roughly 2 box
periods, in spite of which the figure shows that the numerical scheme retains its
third- and sixth-order consistency in time and space, respectively.


\subsection{Initial shearing of isotropic flow}\label{sec:rogers}
To validate the nonlinear terms of the code, the short-term shearing of an initially
isotropic turbulent flow is compared with the classical results of Rogers and
Moin,~\cite{RogersMoin1985} as given in the dataset HOM23 of the AGARD
database,~\cite{AGARD} whose naming notation we use. The initial conditions are
random isotropic fields with a top-hat one-dimensional energy spectrum, as in
Ref.~\onlinecite{Rogallo1981}, adjusted to the same parameters as in
Ref.~\onlinecite{RogersMoin1985}. They all agree well with the reference data, but
the energy dissipation of our DNSes is slightly higher than in the reference cases
after $St \approx 5$ (see later Fig.~\ref{fig:validation1_fine}b), probably because
of the periodic loss of the enstrophy in the remeshing process of Rogallo's Fourier
code.~\cite{LeeKimMoin1990, Rogallo1981} Note that the reference simulations in
Ref.~\onlinecite{RogersMoin1985} were remeshed every $St=2$. Other quantities, such
as the two-point velocity correlation functions, were checked in detail against case
{HOM23U} (Fig.~\ref{fig:validation_corr}). They also agree well, confirming that the
large scales of the present DNS are consistent with those of the three-dimensional
Fourier spectral simulations.

The previous results do not test the effect of dealiasing in $y$, which is applied in 
Ref.~\onlinecite{RogersMoin1985} but not in our case. This is tested in
Fig.~\ref{fig:validation1_fine} by comparing the results of simulating case HOM23U
in its original grid with a much finer grid with $(N_x, N_y, N_z) = (510, 384, 254)$. 
Fig.~\ref{fig:validation1_fine}(a) shows the
temporal evolution of the effective resolutions $\kvec_{\mathrm{max}} \eta$, where 
$\eta(t) = (\nu^3/\diss_V)^{1/4} $ is the instantaneous Kolmogorov scale, and the maximum
effective wavenumbers are $k_{x \mathrm{max}} = \pi/\Delta x$ and $k_{y
\mathrm{max}} = 2.5/\Delta y $. Fig.~\ref{fig:validation1_fine}(b) shows the
evolution of the dissipation rate $\diss_V\equiv \nu \langle \omega_i \omega_i \rangle_V$ 
of the two simulations. The finer grid case has a
slightly larger dissipation rate at around $ 5 \lesssim St \lesssim 10$, but the
agreement is excellent considering that the larger grid is at least twice as fine as
the coarser one. As an added test, the coarser grid was run at CFLs from 0.05 to 0.6,
but the results agree within the thickness of the lines in 
Fig.~\ref{fig:validation1_fine}(b). 

\begin{table}[tb]
\caption{Parameters of present DNS (L11 and M11), CFL=0.6, compared with run2 and
run10 in Ref.~\onlinecite{Pumir1996}. In all cases, $A_{xz}=A_{yz}=1$. The effective resolution
is $\itbold{k} \eta$. The total time to accumulate the statistics is $ST_\mathrm{stat}$. 
$B_2=(3 b_{ij}b_{ij}/2)^{1/2}$ is the second invariant of the Reynolds-stress anisotropy
tensor~\cite{luml78}, $b_{ij} =\bra u_i u_j\ket \bra u_k u_k\ket-\delta_{ij}/3$,
where $\delta_{ij}$ is Kronecker's delta. 
The ratio of energy input and energy dissipation is $\mathcal{P}/\diss$.
The root-mean-squared vorticity magnitude is 
$\omega' = \sqrt{ \bra \omega_i\omega_i\ket }$.
}\vspace{0.5ex}
\centering
\begin{tabular}{c|*{11}{c}}
\hline\hline 
 Run & $Re_z$ & $N_x, N_y, N_z$ & $k_x \eta$ & $k_y \eta$  & $S T_\mathrm{stat}$ & $Re_\lambda$  & $ S^\ast $  & $B_2$ & $-b_{xy}$ &$\mathcal{P}/\diss$ & $ \omega' /S$   \\
\hline 
L11 & 2600 & $62, 96, 62$ & 1.5 & 1.9 & 802 & 52.9 & 6.4  & 0.452 & 0.154 & 0.990 & 6.38 \\
run2 & 2632 & $ 64^3 $ & 1.5 & 1.5  & 210 & 51.6 &6.6 & 0.446 & 0.152 & 1.004  & $6.05$   \\
\hline 
M11 & $8224$ & $ 108, 162, 108$ & 1.1 & 1.40 & 831 &91.0  & 6.9 & 0.454 & 0.142& 0.989 & 10.1\\
run10 & $8225$ & $108^3$& 1.2 & 1.2  & 106 & 83.4 & 7.1 &  0.430 & 0.141& 1.000 & 9.10 \\
\hline\hline 
\end{tabular}\label{table:PumirGualtieriNew}
\end{table}

\subsection{Statistically stationary homogeneous shear turbulence}

Closer to the subject of this paper than any of the previous tests is the long-term
behaviour of small computational boxes. It was shown in Ref.~\onlinecite{Pumir1996} that, under
those circumstances, turbulence grows in size, fills the box, and collapses
intermittently, while reaching a statistically steady state that resembles the
bursting cycle of wall-bounded turbulence~\cite{Jimenez2013nearwall,jim05}. We will
see below that a typical bursting period is $ST_b\approx 25$, so that reasonable
statistics require, depending on the box geometry, running times of the order of
hundreds of box periods. In this section we test the ability of our code to run for
long times by repeating two of the simulations in Ref.~\onlinecite{Pumir1996}.
 
Both simulations run in a cubical box $(A_{xz}=A_{yz}=1)$, and start from
initially turbulent conditions. In each case, we accumulate statistics for $S
t\approx 800$, after discarding the initial  $St \approx 30$. The
energy input by action of the the shear on the stress, $\mathcal{P} \equiv -S\uvp$, 
balances the dissipation rate within 1\%.

Table~\ref{table:PumirGualtieriNew} compares our two simulations (L11 and M11) with
those in Ref.~\onlinecite{Pumir1996}. It includes both small- and large-scale quantities, which
agree well. It is probably significant that the quantities that depend on the small
scales, such as $Re_\lambda$ or $\omega^\prime/S$, tend to be somewhat
higher in our simulations than in Ref.~\onlinecite{Pumir1996}. This is consistent with the loss
of enstrophy in the spectral code due to remeshing, although the difference is too
small to decide whether the reason in this particular case is numerical or
statistical.


\bibliography{./database}

\end{document}